\documentclass[
reprint,
superscriptaddress,
nofootinbib,
amsmath,amssymb,
 aps,
prd,
floatfix,
author-numerical,
]{revtex4-2}

\usepackage{graphicx}% Include figure files
\usepackage{dcolumn}% Align table columns on decimal point
\usepackage{bm}% bold math
\usepackage[hidelinks]{hyperref}% add hypertext capabilities
\usepackage[capitalise]{cleveref}
\usepackage{gensymb}
\usepackage{tabularx}
\usepackage{booktabs}
\usepackage{verbatim}
\usepackage{bbold}
\usepackage{braket}
\usepackage{aas_macros}
\usepackage{xcolor}
\usepackage{comment}

\newcommand*{\Comb}[2]{{}^{#1}C_{#2}}%
\newcolumntype{P}[1]{>{\centering\arraybackslash}p{#1}}
\newcommand\comCGL[1]{{\color{red} CGL[ #1 ]}}
\begin{document}

\preprint{APS/123-QED}

\title{On the spatial distribution of the Large-Scale structure:\\ An Unsupervised search for Parity Violation}% Force line breaks with \\
%\thanks{A footnote to the article title}%

\author{Samuel Hewson}
\affiliation{Astrophysics Group, Cavendish Laboratory, University of Cambridge}
\affiliation{St John's College, St John's Street, Cambridge, CB2 1TP, United Kingdom}

\author{Will Handley}
%\email{w.handley@cam.ac.uk}
\affiliation{Astrophysics Group, Cavendish Laboratory, University of Cambridge}
\affiliation{Kavli Institute for Cosmology, University of Cambridge}
\affiliation{Gonville \& Caius College, Trinity Street, Cambridge, CB2 1TA, United Kingdom}

\author{Christopher G. Lester}
\affiliation{High Energy Physics Group, Cavendish Laboratory, University of Cambridge}

%\collaboration{MUSO Collaboration}%\noaffiliation

%\collaboration{CLEO Collaboration}%\noaffiliation

\date{\today}% It is always \today, today,
             %  but any date may be explicitly specified

\begin{abstract}
We use machine learning methods to search for parity violations in the Large-Scale Structure (LSS) of the Universe, motivated by recent claims of chirality detection using the 4-Point Correlation Function (4PCF), which would suggest new physics during the epoch of inflation. This work seeks to reproduce these claims using methods originating from high energy collider analyses. Our machine learning methods optimise some underlying parity odd function of the data, and use it to evaluate the parity odd fraction. We demonstrate the effectiveness and suitability of these methods and then apply them to the Baryon Oscillation Spectroscopic Survey (BOSS) catalogue. No strong evidence for parity violation is detected.
\end{abstract}

%\keywords{Suggested keywords}%Use showkeys class option if keyword
                              %display desired
\maketitle

%\tableofcontents

\section{\label{sec:level1}Introduction}

A parity transformation acts to spatially invert a system and parity has, for a long time, been a defining symmetry of the Standard Model of particle physics. Over the past 30 years, progress has been made in investigating whether parity has a role to play in cosmology. Until very recently most of that work was focused on Cosmic Microwave Background (CMB) Polarisation~\citep{PhysRevLett.83.1506.lue, Kamionkowski_2011, Shiraishi_2016, PhysRevLett.125.221301.komatsu, philcox2023cmb, philcox2024testing}, or on Gravitational Waves~\citep{Saito_2007, Yunes_2010, Wang_2013, Orlando_2021, jenks2023parameterized}.

%Possible Chern-Simons discussion - also Hou et al

In recent years, following the proposal of \citet{cahn2021test} the Large-Scale Structure (LSS) of the Universe has been probed for parity violations using the 4-Point Correlation Function (4PCF). So far these studies have used the Baryon Oscillation Spectroscopic Survey (BOSS)~\citep{10.1093/mnras/stx721} of the Sloan Digital Sky Survey (SDSS)-III~\citep{2011AJ....142...72E, Dawson_2013} and have reported findings of parity violation at the $8.1\sigma$~\citep{philcox2021detection}, $7.1\sigma$~\citep{Hou_2023} and $2.9\sigma$-levels~\citep{Philcox_2022}.

\begin{figure}
    \centering
    \includegraphics{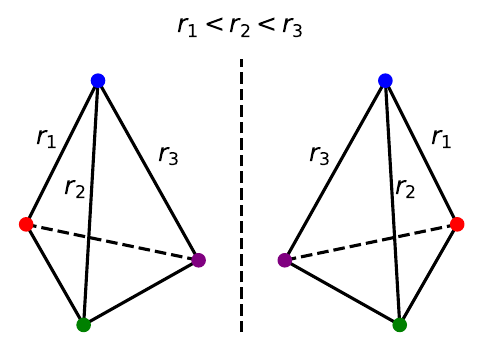}
    \includegraphics{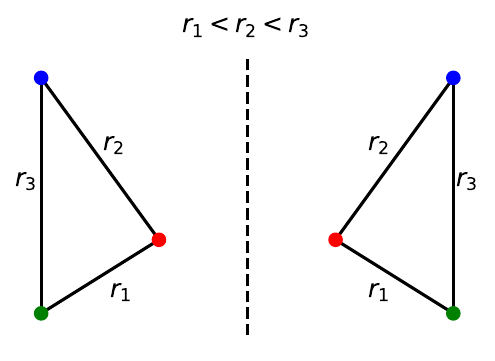}
    \caption{\textbf{Top:} A tetrahedron is the lowest order polygon subject to parity violation in three dimensions, therefore the 4PCF is the lowest order $N$-point correlation function which can be probed for parity violations. \textbf{Bottom:} three points can be parity violating in two dimensions, but in three dimensions the 2d parity operation becomes equivalent to a 180\degree\ rotation. Consider rotating into the page about the reflection axis. Figure adapted from \citet{taylor2023}.}
    \label{fig:tetrahedron_mirror}
\end{figure}

In three dimensions the lowest order polyhedron subject to parity asymmetry is the tetrahedron, as shown in \cref{fig:tetrahedron_mirror}. The 4PCF is, therefore, the lowest order $N$-point correlation function which can be probed for parity violations~\citep{cahn2021test}. Tetrahedra are constructed with galaxies at each of the four vertices, and the relative positions are used to calculate the 4PCF. This method facilitates searches across groups of four galaxies but is, by design, constrained to only search for parity violations at the four-point level. This limits the type and order of parity violations that could be detected. It is possible to consider higher $N$-point correlation functions, but these quickly become intractable. On top of this there could be parity violations which manifest through multiple orders of correlation, which would be undetectable.

A second drawback of the method is the apparent difficulty in quantifying the significance of the detection. \citet{Philcox_2022} performed both a rank test and a $\chi^2$ test, and compared the results to 2048 MultiDarkPATCHY simulations. These simulations are calibrated to match the BOSS two- and three-point clustering statistics~\citep{Rodr_guez_Torres_2016, Kitaura_2016}, not the 4-point, which could lead to an underestimate of the uncertainty in the 4PCF calculations. \citet{Hou_2023} quantify the significance by calculating the uncertainty in the covariance matrix, this could be influenced by observational effects and instrumental systematics ~\citep{Hou_2023}, and assumes that the 4PCF follows a Gaussian distribution~\citep{2022PhRvD.106d3515H}. In fact, recent work by  \citet{krolewski2024evidenceparityviolationboss} shows that correctly accounting for 8-point contributions reduces the significance of detection down to between $0\ \mathrm{and}\ 2.5\sigma$.

The aim of this paper is to demonstrate that more general methods can be used to search for parity violations in the LSS. Such methods do not have to be specifically 4-point, or even $N$-point, but rather consider the data in spatially localised regions and search for underlying parity violations across all $N$-point correlations. A more specialised method will than be implemented for the 4-point case, constructed with no knowledge of a 4PCF, and implemented to see if the results of \citet{Hou_2023} and \citet{Philcox_2022} can be reproduced.

Similar methods to the ones we applied have also been proposed and tested with CNNs and Neural Field Scattering Transforms (NSFT) \citep{taylor2023, craigie2024unsupervisedsearchescosmologicalparity}.  With the recent data release from the Dark Energy Spectroscopic Instrument (DESI)~\citep{2016arXiv161100036D, https://doi.org/10.5281/zenodo.7964161} and the upcoming releases from EUCLID~\citep{2011_EUCLID_def, 2020_Euclid} and the Nancy Grace Roman Space Telescope~\citep{2015Roman, 2019Roman}, the methods proposed her will be useful in future exploration of the LSS.
\vspace{-0.4em}

\section{Theory and Methods}

\subsection{Theory of Parity}

In Quantum Mechanics the Parity Operator, $\hat{P}$ is defined to spatially invert the wavefunction of a particle, such that a wavefunction $\Psi(x)$ is Parity-even if $\Psi(\hat{P}x)\ =\ \Psi(x)$ and Parity-odd if $\Psi(\hat{P}x)\ =\ -\Psi(x)$. The importance of parity pervades throughout physics, most notably in the Standard Model of particle physics, where parity is a symmetry of both Quantum Electrodynamic (QED) and Quantum Chromodynamic (QCD) interactions, but is violated by Weak interactions~\citep{PhysRev.104.254-weaktheory, PhysRev.105.1413-weakexp}. The creation of the present-day matter-antimatter asymmetry would require baryogenesis processes that violate charge and parity conservation~\citep{Sakharov:1967dj, Carroll_1998}, but since gravity is parity symmetric~\citep{Hilbert1915}, all cosmological correlators should be too. This means that parity asymmetry in the LSS would require new physics in the epoch of inflation.

\subsection{Method of Lester and Tombs}

Current theoretical models predict parity violating mechanisms which are not observable at the Large Hadron Collider (LHC)~\citep{Lester_2019}, and accordingly the LHC has not been used to search for these. Parity may however be violated by unknown means, which could be detectable in LHC deposit data. Motivated by this possibility, \citet{lester2022using} demonstrated that machine learning can be used to detect parity violating functions within datasets without any required knowledge of what is causing the violation. By training an image classification Convolutional Neural Network (CNN) and rewarding it for detecting asymmetry, a measure of the parity of the image dataset can be determined. 
To demonstrate this method, \citet{lester2022using} applied their algorithm to the MNIST data as well as more general symmetries~\citep{Tombs_2022}. The success of the algorithm on the MNIST data also displays the generality of the method, and its applicability outside particle physics. To further demonstrate the general success of the \citet{lester2022using} method, a Master's project was run in 2022/3 aiming to use the method to test for parity violations in snails~\citep{lester_gibbon2022}. The population was known to be parity violating as most snail species have a strong preference for the handedness of their shells' spirals. It was successfully shown that the unsupervised learning method could detect parity violations within the images.

\subsubsection{Algorithm}

To quantify any parity violations existing in a sample, consider some function $g(x)$, where $g$ represents the outputs of the CNN with inputs $x$. Then construct $f(x)$ to be the difference between the outputs from $x$, and the parity transform of $x$: 
\begin{equation}\label{f}
    f(x) = g(x) - g(\hat{P}x)
\end{equation}
from which it is immediately clear that $f(\hat{P}x) = -f(x)$. In fact further symmetries of the system can be specified: let ${S = \{S_1, S_2, ..., S_i\}}$ be the set of all the symmetries of the system. Then $f(x)$ can be defined as the sum over $S_i$ of the difference between the outputs from the original and parity transformed inputs as demonstrated in \cref{general_parity_f}:
\begin{equation}\label{general_parity_f}
        f(x) = \sum_{i}{g(\hat{S_i}x) - g(\hat{S_i}\hat{P}x)}
\end{equation}
\cref{f180_rot} is an example of this where $f(x)$ has been constructed to obey invariance under 180\degree\ rotations, represented by $\hat{R}_{180}$.
\begin{equation}\label{f180_rot}
    f(x) = g(x) + g(\hat{R}_{180}x) - g(\hat{P}x) + g(\hat{R}_{180}\hat{P}x)
\end{equation}
This creates a function for which $f(x) = f(\hat{R}_{180}x)$, enforcing 180\degree\ rotational symmetry and removing any idea of a `top' from the images.
Accounting for symmetries in $f$ means they do not need to be accounted for in the subnetwork $g$, reducing the processing required on the inputs. Further symmetries are enforced by max-pooling layers within the network.

\cref{f} will output zero for a parity even function $g(x)$, and the network can be constructed to return outputs $g$ such that the difference of $f$ from zero is maximised. For each batch $B = \{x_1, x_2, ..., x_{\lvert B \rvert} \}$, the mean value of $f$ is calculated.
\begin{equation}\label{mean_f}
    \mu_B = \frac{1}{\lvert B \rvert}\sum_{x \in B} f(x)
\end{equation}
The loss function is defined 
\begin{equation}\label{loss_func}
    \mathcal{L}_B = -\mu_B /\sigma_B,
\end{equation} 
where
\begin{equation}
    \sigma_\mathrm{B}^2 = \frac{1}{\lvert B \rvert}\left(\sum_{x\in B}[f(x)]^2\right) - \mu_B^2
\end{equation}
is the variance of $f$ over batches, which ensures that the network will not trivially increase the loss under $f(x) \mapsto \lambda f(x)$ for $\lambda > 1$.  

After training the neural net, the function $f(x)$ can be evaluated on the testing data. If the mean value of $f$ on that dataset differs from zero by a statistically significant amount, some parity-violating feature in the data has been detected. We compute the fraction of the outputs for which $f(x) > 0$. This `positive fraction' quantifies the proportion of the inputs that exist in a parity violating way, which means achiral datasets are expected to output 50\%, and deviations from this imply a parity violating dataset. 

It is important to note here that when searching for a violation of some symmetry then a function $f$ can provide evidence if \textbf{both}:
\begin{enumerate}
    \item $f$ has the correct symmetry properties to permit the discovery, and
    \item $f$ yields statistically significant outputs when evaluated on independent datasets that it has not seen before and that were not used in its construction
\end{enumerate}
Here we have constructed $f$ by hand to be parity-odd, which satisfies the first requirement. The second requirement is demonstrated in \cref{sec:Valid}.
From these principles it follows that the most important step here is the construction of $f$. This method could, in principle, work without any training. The training step just acts to increase the likelihood of a discovery by optimising, but not changing the form, of $f$.
This also means that the net has the benefit of never creating a false positive from overtraining. If there is a positive detection, the data must inherently contain the parity violation specified by the function $f$, regardless of whether that $f$ was obtained by overtraining.

\section{Dataset}
\label{sec:Data}
\begin{figure}
    \centering
    \includegraphics[width=\linewidth]{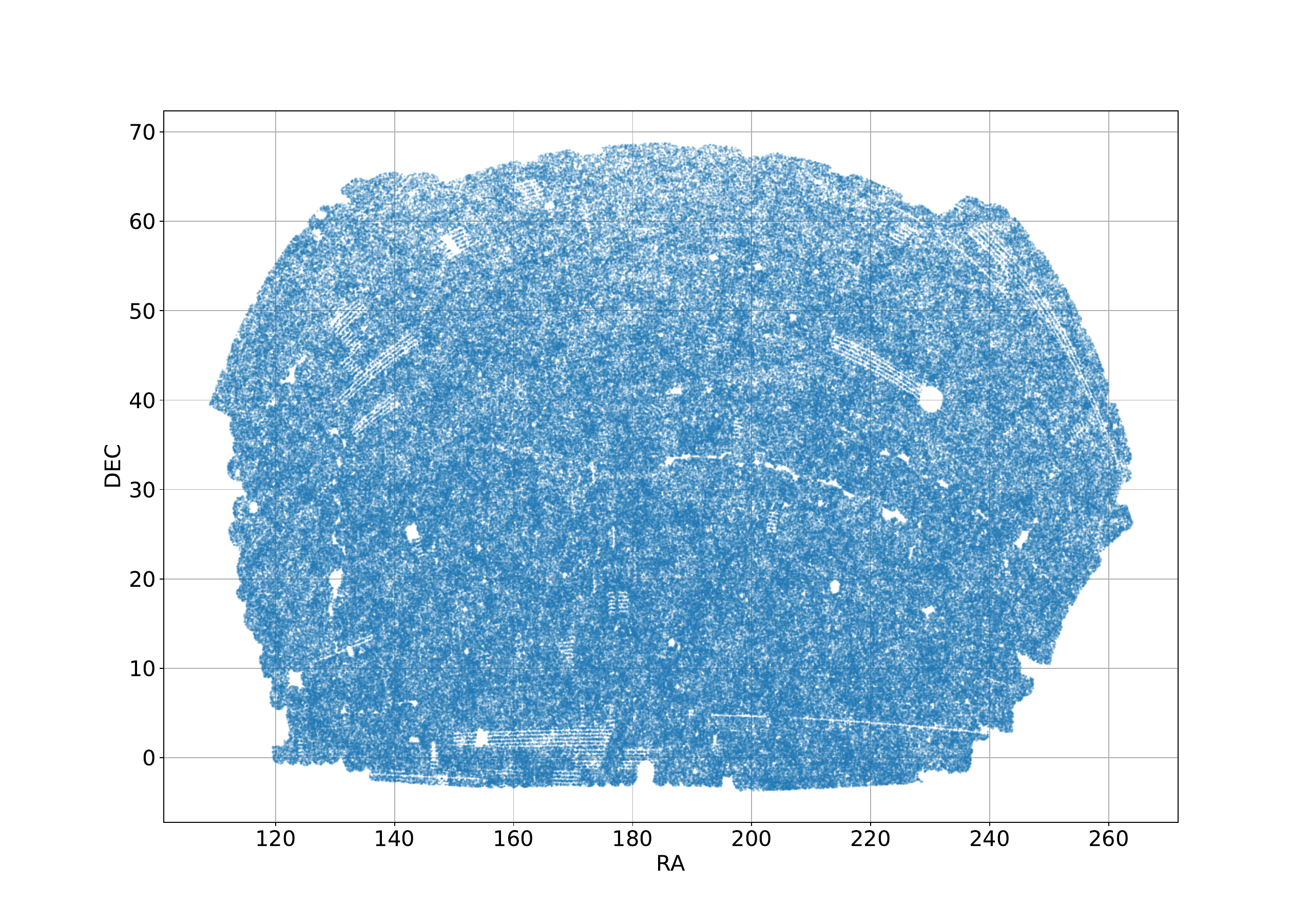}
    \caption{The North Galactic Cap of the CMASS catalogue.}
    \label{fig:cmass_NGC}
\end{figure}

Following from \citet{Hou_2023} and \citet{Philcox_2022}, we use data release 12 (DR12) of the BOSS catalogue, which contains LOWZ and CMASS samples, each being split into the North Galactic Cap (NGC) and South Galactic Cap (SGC). The target selection algorithm results in samples of primarily Luminous Red Galaxies (LRGs), with CMASS further specified to select objects of uniform redshift~\citep{2016_REID_BOSS}. This leads to a roughly mass-limited sample down to a stellar mass $M \sim 10^{11.3}h^{-1}M_\odot$~\citep{marston_mNRAS_2013, 10.1093/mnras/stw1080, 10.1093/mnras/stw117, Bundy_2017}. A redshift cutoff of $0.43 < z < 0.7$ is applied, which raises the purity of the CMASS sample by adding more LRGs~\citep{Hou_2023}.
For this investigation, the CMASS NGC was used, and the catalogue of galaxies was treated as a field of points in three-dimensional space, a 2d projection of this is shown in \cref{fig:cmass_NGC}. The BOSS dataset contains data in the form of right ascension, declination and redshift. This was converted into spatial coordinates using the fiducial cosmology specified by the BOSS collaboration~\citep{Alam_2017}. Namely, a flat $\Lambda$CDM model, with matter density $\Omega_\mathrm{m} = 0.31$, Hubble constant $h \equiv H_0/(100\text{km\ s}^{-1}\ \text{Mpc}^{-1}) = 0.676$, baryon density $\Omega_\mathrm{b}h^2 = 0.022$, fluctuation amplitude $\sigma_8 = 0.8$ and spectral tilt $n_\mathrm{s} = 0.97$. From this, the question can be asked: Is the spatial distribution of the points in the field parity violating?
More specifically, is the spatial distribution of the points in the field parity violating on a given scale? The scale of the parity violation is specified to match the work of \citet{Hou_2023}. This means for the geometric analyses, groups are constructed such that points within a group are separated by a distance, ranging from $20h^{-1}$Mpc to $160h^{-1}$Mpc.

To each galaxy a weight was applied:
\begin{equation}
    w = w_\mathrm{fkp}w_\mathrm{sys}(w_\mathrm{rf} + w_\mathrm{cp} - 1)
\end{equation}
where $w_\mathrm{rf}$ is the redshift failure weight and $w_\mathrm{cp}$ the fibre collision weight; these are additive weights with default value of unity. $w_\mathrm{sys}$ is the systematic weight, which accounts for stellar density and seeing~\citep{Anderson_2012_weights}, and the FKP weight is defined as ${w_\mathrm{fkp} = [1+n(z)P_0]^{-1}}$~\citep{Feldman_1994_fkp}, where $n(z)$ is the weighted number density as a function of redshift, and $P_0 = 10^4[h^{-1}\text{Mpc}]^{-3}$.

\section{Validating the Methods}\label{sec:Valid}

To begin, in \cref{sec:2d_conv} we use the CNN architecture of \citet{lester2022using} to examine a two-dimensional projection of the data. This is followed in \cref{sec:3d_CNN} by the addition of colour channels to the images, with the aim of representing three-dimensional information via colour. Finally, we investigate more geometric approaches in \cref{sec:angle}, which have the drawback of being constrained to some order of polyhedron, much like $N$-point correlation functions, but consequently allow direct comparison with the 4PCF.

\subsection{2D Convolution}\label{sec:2d_conv}

We begin by using the same architecture as \citet{lester2022using}: 2d convolutional layers, with image inputs.
For our investigations, the data can be treated as a field of points, from which we constructed images to represent localised regions of the field.
To ensure that this method is suitable for detecting parity violations in image data, we ran a number of tests using simulated data. These not only verify that the network is able to detect parity asymmetric data, but also help to determine a level of significance i.e.\ what net fraction of the data must contain a parity violating object to result in a non-zero detection.\ 
The choice of how to represent galaxies in the images is somewhat arbitrary, and two possible examples are shown in \cref{fig:Demo_images}. After testing multiple methods (\cref{appendix:image_sizes}), we decided to use PNG images generated from scatter plots of data points. An outline of this method follows: 

\begin{figure}
    \centering
    \includegraphics[width=\linewidth]{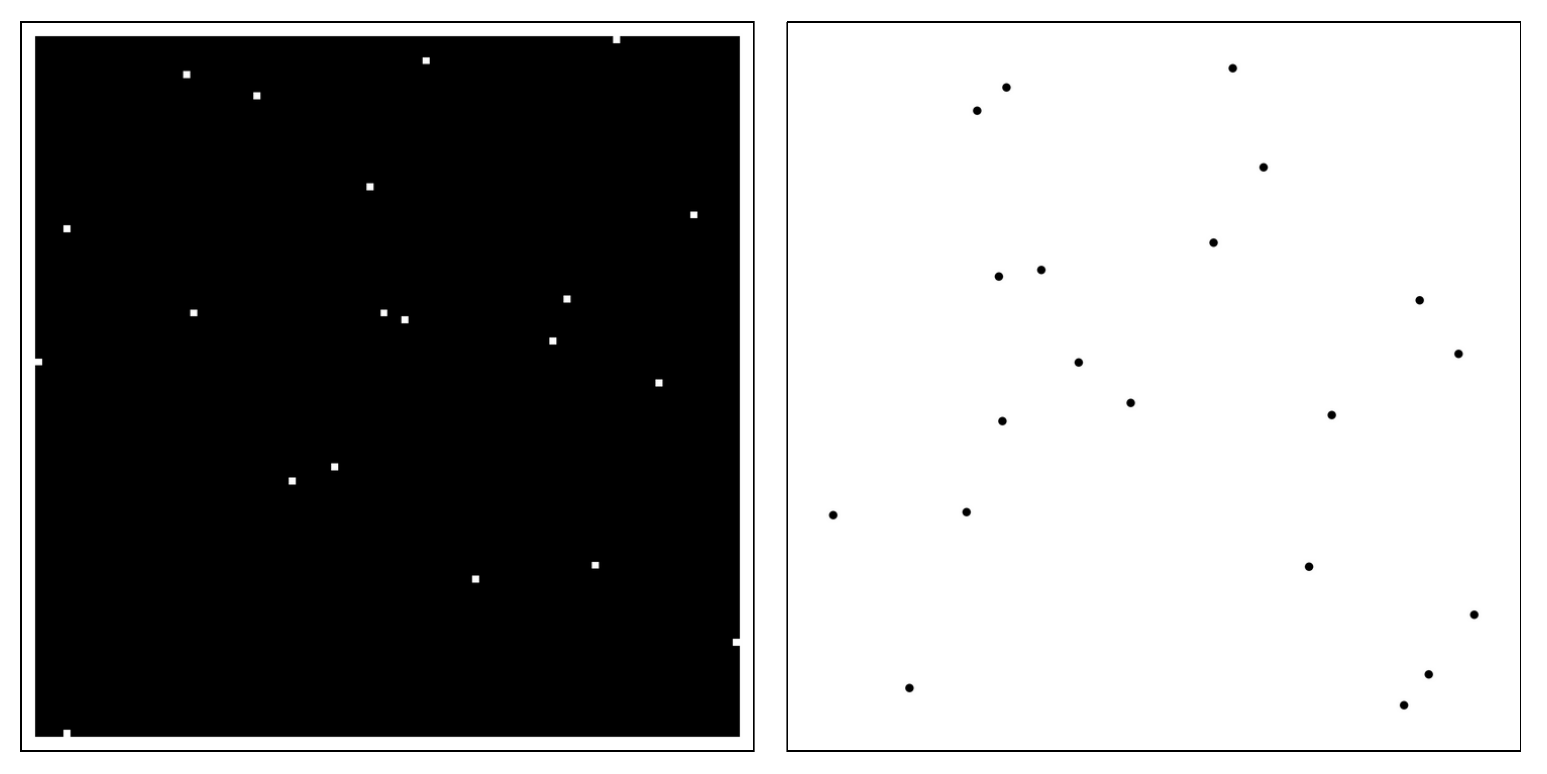}
    \caption{Two demonstration images generated from random data. \textbf{Left:} An image where each `galaxy' has its localised pixels colourised. \textbf{Right:} An image where each `galaxy' is represented by a scatter point.}
    \label{fig:Demo_images}
\end{figure}

To begin, a field of random data points was generated, and image samples created in the following manner:
\begin{enumerate}
    \item Sample a random point from the dataset, $O$
    \item Generate a square of side length $l$, centred on $O$
    \item Rotate the square by a random angle, $\theta \in [0, 2\pi]$
    \item Get all the data points lying within the rotated square
    \item Create an image of the square of data points
\end{enumerate}

\begin{figure}
    \centering
    \includegraphics[width=0.8\linewidth]{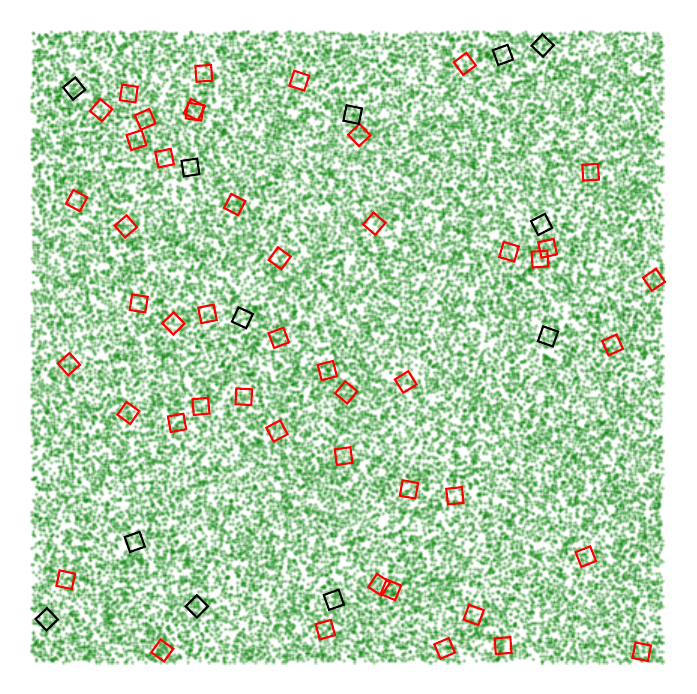}
    \caption{A demonstration of the image sampling technique, on a background random field of 500,000 data points, with 12 testing samples shown in black and 48 training samples shown in red.}
    \label{fig:demo_random_imgsample}
\end{figure}

This ensures that all the images are the same size, 64x64 pixels, following the work of \citet{lester_gibbon2022}. Testing images are generated first, and then training ones with the added condition that these images may not overlap with the testing set, to avoid overfitting. An exaggerated sample is displayed in \cref{fig:demo_random_imgsample} to clearly show the method of sampling squares.

\subsubsection{Testing method with toy data}

The toy data tests were constructed on the same scale as the real images would be. As outlined in \cref{sec:results} images covering 0.5\degree x 0.5\degree\ were chosen, giving an average of 22.3 galaxies per image. For the toy models we generated 12,000 images, split into 9,600 training and 2,400 testing, with each image averaging 22.3 data points.

\begin{figure}
    \centering
    \includegraphics[width=\linewidth]{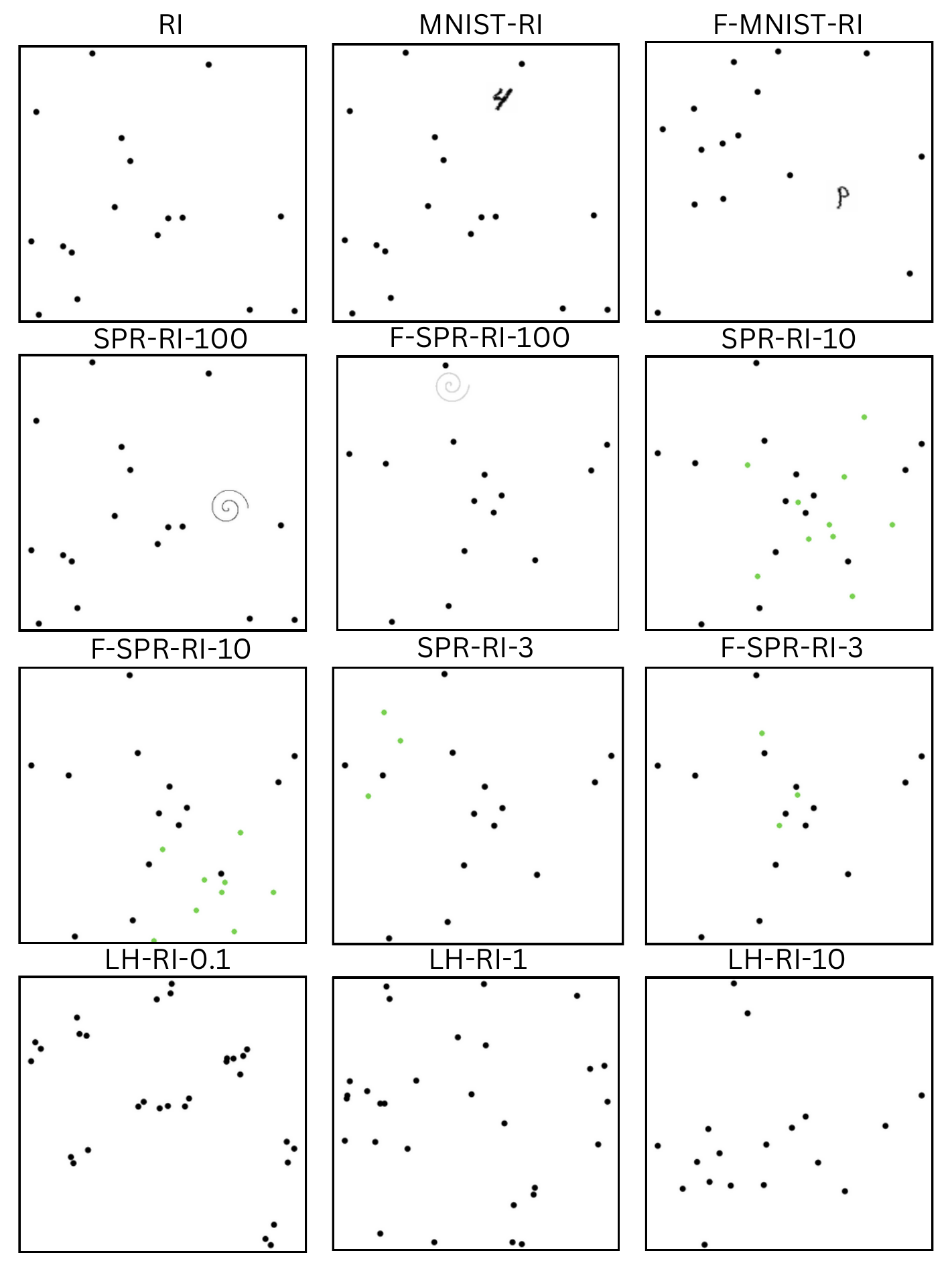}
    \caption{A sample of toy images labelled by their data codes from \cref{table:CNN_datacodes}. For the purpose of demonstration, smaller spirals that have been overlaid have been coloured green. During testing, all images where in black and white. }
    \label{fig:Demo_samples}
\end{figure}

\begin{comment}
\comCGL{Thank you sam for making the footnote appear as a footnote. Alas, it somehow seems partially clipped. The end is missing? }
\end{comment}
A number of toy datasets were curated in order to investigate the effectiveness of the method under different scenarios. The critical datasets are outlined with a description and purpose in \cref{table:CNN_datacodes}, with example images displayed in \cref{fig:Demo_samples}. Each dataset has an associated control dataset, which ensures the function of the net. A control group is generated by iterating through the real group, and flipping each image with a 50\% probability. For any large parity odd dataset this symmetrisation process always creates an achiral dataset, which can act as a control.\footnote{A right-handed figure drawn on a flat piece of tracing paper becomes left-handed if you are allowed to turn the tracing paper over and view it from the back. The handedness of two-dimensional objects are therefore only well-defined if the surfaces on which they live are orientable or are oriented in some special way(s).  The toy datasets in \cref{table:CNN_datacodes} are all intrinsically two-dimensional and so, for the reason just given, any of these datasets which  violate parity only do so on account of the embedded objects being implicitly pasted the `inside' of the celestial sphere, centered on earth, from where they are viewed.  Such geocentric parity violation is, of course, entirely unlike the sorts of large scale parity violation one might want to see in the real universe. Any large scale parity violation will (presumably) not rely on the earth being at a special place in the universe.  The geocentric nature of the parity violation in these toy datasets does not prevent them being used for the purposes they were created, however.  Furthermore,  this geocentricity is completely removed in the parity violating datasets introduced later in \cref{table:3d_cnn_datacodes}.  These later toy datasets contain projections of intrinsically three-dimensional (parity violating) structures without any preferred orientations, and so are representative of a sort of parity violation that one might hope to see in reality.} Each data type was created from the same randomly generated background field, repeated over 10 different background fields to create 10 variations of each data type. The average `positive fraction' over the 10 repeats is listed in \cref{table:CNN_datacodes}. 

\begin{table*}
    \begin{tabular}{  p{2.35cm}  p{5.5cm}  p{4.9cm} P{2.0cm} P{2.0cm}}
        \toprule
\textbf{Data code}      
& \textbf{Description}   
& \textbf{Purpose} 
& \textbf{Detection}
& \textbf{Control}\\\midrule
RI
& Random Generated Image Data      
& Check sampling method isn't introducing parity violation
& $49.9\pm0.3\%$ 
& $49.8\pm0.3\%$ \\\hline
MNIST-RI     
& RI with random MNIST image overlaid in random position                     
& Check detection of complex parity violation  
& $87.5\pm0.3\%$ 
& $50.3\pm0.3\%$ \\\hline
F-MNIST-RI      
& MNIST-RI, but each MNIST image has a  50\% chance of being flipped vertically before inserting
& Check method of insertion is not causing spurious detection 
& $50.0\pm0.3\%$ 
& $50.4\pm0.3\%$ \\\hline
SPR-RI-100 
&RI with 100 point spiral inserted in random position and random radius    
& Is parity violation in the same form as the data detected 
& $98.2\pm0.1\%$ 
& $49.5\pm0.3\%$ \\\hline
F-SPR-RI-100
&SPR-RI-100, but each spiral has a  50\% chance of being flipped vertically before inserting
&Check method of insertion is not causing spurious detection 
& $49.6\pm0.3\%$ 
& $49.5\pm0.3\%$ \\\hline
SPR-RI-10
&RI with 10 point spiral inserted in random position and random radius      
& Is parity violation detectable on a more realistic scale
& $97.9\pm0.1\%$ 
& $50.1\pm0.3\%$ \\\hline
F-SPR-RI-10
&SPR-RI-10, but each spiral has a  50\% chance of being flipped vertically before inserting
&Check method of insertion is not causing spurious detection 
& $49.8\pm0.1\%$ 
& $51.6\pm0.3\%$ \\\hline
SPR-RI-3
&RI with 3 point spirals (triangles) inserted in random position and of random radius      
& Is parity violation detectable on the smallest scales
& $93.5\pm0.2\%$ 
& $51.1\pm0.3\%$ \\\hline
F-SPR-RI-3
&SPR-RI-3, but each spiral has a  50\% chance of being flipped vertically before inserting
&Check method of insertion is not causing spurious detection 
& $51.9\pm0.3\%$ 
& $49.2\pm0.3\%$ \\\hline
LH-RI-0.1
& RI where every point has 2 extra points generated close to it forcing a parity odd `left-handed' triangle on the scale of 1 tenth of the average inter-point distance
& Can other types of parity violation be detected 
& $74.6\pm0.3\%$ 
& $49.6\pm0.3\%$ \\\hline
LH-RI-1
& LH-RI-0.1 with extra points on the scale of the average inter-point distance
& Can more realistic types of LH-RI-0.1 parity violation be detected 
& $67.6\pm0.3\%$ 
& $49.2\pm0.3\%$ \\\hline
LH-RI-10
& LH-RI-0.1 with extra points on the scale of ten times the average inter-point distance
& Can more realistic types of LH-RI-0.1 parity violation be detected
& $52.3\pm0.3\%$ 
& $49.13\pm0.3\%$ \\\hline
SPR-RI-5-1\%
&RI with 5 point spiral inserted in random position and random radius in 1\% of images      
& How sensitive is the detection
& $57.6\pm0.3\%$ 
& $49.8\pm0.3\%$ \\\hline
SPR-RI-5-0.1\%
&RI with 5 point spiral inserted in random position and random radius in 0.1\% of images      
& How sensitive is the detection
& $47.6\pm0.3\%$ 
& $49.5\pm0.3\%$ \\\hline
SPR-RI-5-0.01\%
&RI with 5 point spiral inserted in random position and random radius in 0.01\% of images      
& How sensitive is the detection
& $48.0\pm0.3\%$ 
& $48.8\pm0.3\%$ \\\bottomrule
    \end{tabular}
    \caption{A summary of the crucial image based toy datasets. Each dataset is given a code and a description and has its testing purpose outlined. The detection column contains the positive fraction detection, and the control column shows the positive fraction for the control group for each test.}
    \label{table:CNN_datacodes}
\end{table*}

\subsubsection{Summary of tests}
The results of the toy data samples demonstrate that the net does not find parity violations where there are none and that it does detect parity violations where they exist on large enough scales. Crucially, it also shows that the method of sampling is not creating spurious detection. The ability to detect varying types of complex parity violation suggests that this method should be successful in finding any such violations in the real data.

These tests also give insight into the sensitivity of the method: having just 0.1\%  of the images contain parity violation is enough for it to be successfully detected.
Perhaps the most powerful tests are \textbf{SPR-RI-3} and \textbf{LH-RI-10}, which show that images without any visible asymmetry, can be detected by the net.

\subsection{2D Convolution with colour}\label{sec:3d_CNN}

The 2d CNN is constrained by its loss of 3 dimensional information, meaning not all forms of 3d parity violation are detectable. A first attempt to resolve this is made by introducing colour to the images. By normalising the axis into the page and translating it to pixel colour, it is possible to add extra information to the two-dimensional images. With this modification each image represents a voxel of the space displayed as a 2d projection, with the third dimensional information encoded in the colour of a given point, this creates a colour CNN (cCNN).

As in the previous case, we generate a random field of points, now in 3d space. Images are constructed following the same process outlined in \cref{sec:2d_conv}, with the additional step of using z values and weights to colour the pixels. Under this construction, each pixel has two degrees of freedom we need to express. The first is the distance of the galaxy and the second is its weight. To represent these degrees of freedom we assign RGB values to the pixels in the following manner:

\begin{equation}
    (r, g, b)_{jk} = \sum_{i \in jk}{w_i\cdot\bm{c}(z_i)}
\end{equation}

\begin{equation}
    \bm{c}(z) = z \cdot \bm{R} + (1 - z) \cdot \bm{B} 
\end{equation}

\begin{figure}
    \centering
    \includegraphics[width=\linewidth]{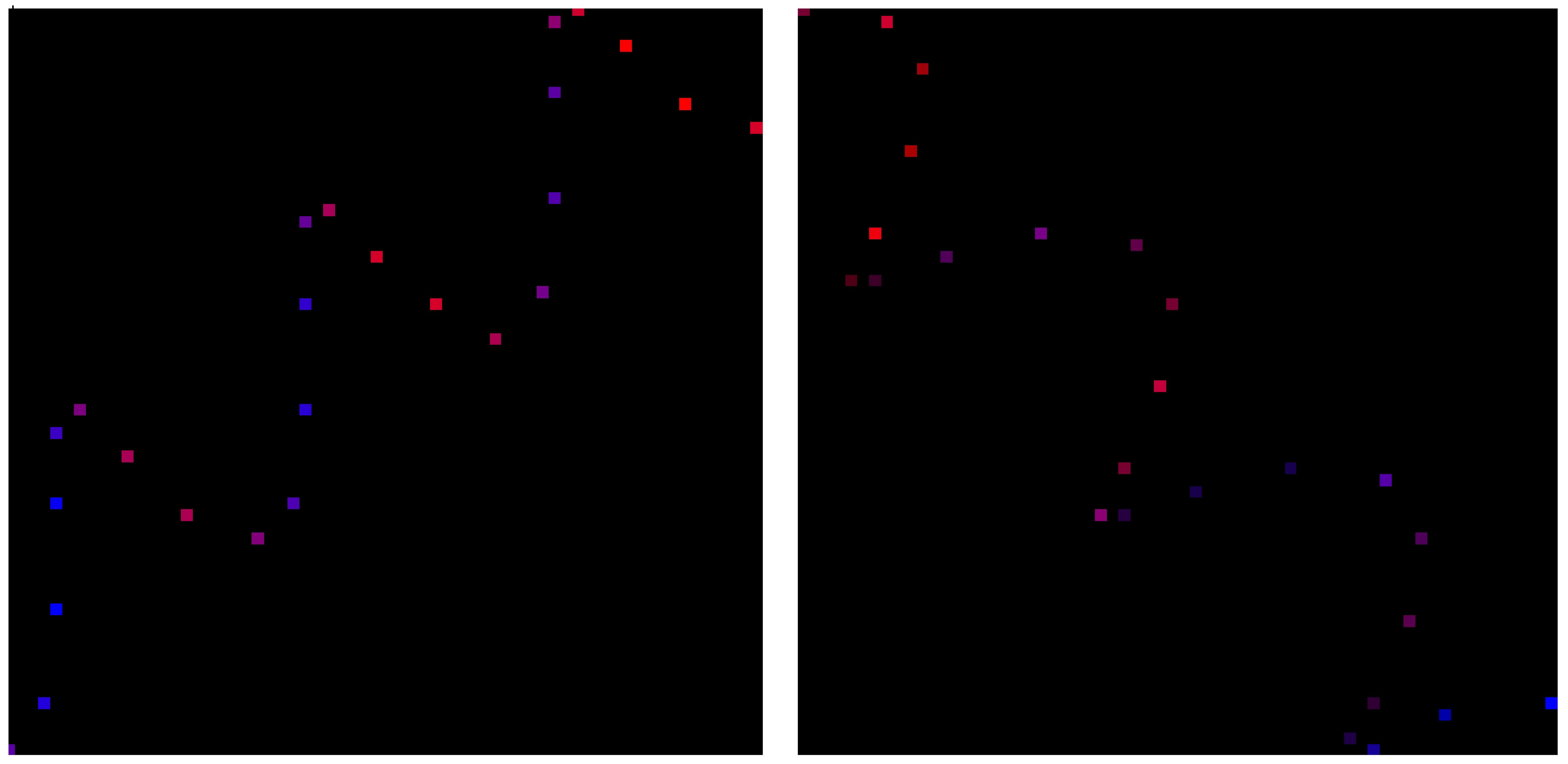}
    \caption{Sample images containing helices. The colouration creates a parity violating dataset. \textbf{Left:} all galaxies have weight 1.0, and therefore are all the same intensity. \textbf{Right:} Galaxies have a randomly assigned weight from (0.25, 0.5, 0.75), this affects intensity.}
    \label{fig:demo_helices}
\end{figure}

Where $w_i$ is the weight of the $i$th galaxy in pixel $jk$ and $\bm{R} = (1, 0, 0)$ and $\bm{B} = (0, 0, 1)$. In this way the distance correlates to the ratio between the red and blue channels, and the weight is represented by the intensity of the pixel. The background is set to $(0, 0, 0)$, so the net is looking at $\mathcal{O}(1)$ deviations from 0. A demonstration of this is visible in \cref{fig:demo_helices}. 

To account for the slight issues that arise from the fact that some the weights are larger than 1.0 and the fact that the clipping of values to fit in the integer range (0, 255) will cause some rounding issues, we also create an array dataset. This dataset contains the same 3 colour channel arrays of 64x64 pixels as would be extracted from the images, but contains slightly improved accuracy values. This array can then easily be manipulated in the same way as the image data.

Under this construction we use the following identity to simplify the parity operator, to just consider an $x$ flip:
\begin{equation}
    \hat{P} \equiv \hat{P}_x\hat{P}_y\hat{P}_z = \hat{P}_{i\in \{x,y,z\}}\hat{R}
\end{equation}

We then enforce rotational invariance in the 2d plane, so for inputs $\alpha$ construct $f(\alpha)$ in the following way:
\begin{equation}
    f(\alpha) =\sum_{\hat{R}_i}\ \left[g(\hat{R}_i\alpha) - g(\hat{R}_i\hat{P}_x\alpha) \right], \label{eq:comb_f}
\end{equation}
where $\hat{R}_i \in [\hat{R}_{0}, \hat{R}_{90}, \hat{R}_{180}, \hat{R}_{270}]$. For this setup then, we can use the same net as we used for the black and white images.

\subsubsection{Testing method with Toy Data}

\begin{table*}
    \begin{tabular}{ p{2.0cm}  p{5.2cm}  p{4.0cm} P{1.9cm} P{1.95cm} P{1.95cm}}
        \toprule
\textbf{Data code}      
& \textbf{Description}   
& \textbf{Purpose} 
& \textbf{Detection}
& \textbf{Control}
& \textbf{Uncoloured} \\\midrule
RCI
& Random Generated Image Data      
& Check sampling method isn't introducing parity violation
& $49.8\pm0.3\%$ 
& $50.3\pm0.3\%$ 
& $49.9\pm0.3\%$ \\\hline
HE-RCI     
& RCI with 1 left-handed helix inserted into the dataset for every 100 random points                    
& Check detection of complex parity violation  
& $57.5\pm0.3\%$ 
& $50.2\pm0.3\%$ 
& $49.8\pm0.3\%$ \\\hline
F-HE-RCI      
& HE-RCI, but each helix has a  50\% chance of being left- or right-handed 
& Check method of insertion is not causing spurious detection 
& $50.5\pm0.3\%$ 
& $50.4\pm0.3\%$ 
& $50.2\pm0.3\%$ \\\hline
PV-RCI  
& Parity violating groups of 4 points randomly oriented and distributed. Local separation $\ll$ average interpoint distance.  
& Can parity violations be detected where the 2d dataset does not detect
& $87.2\pm0.1\%$ 
& $49.6\pm0.3\%$ 
& $50.8\pm0.3\%$ \\\bottomrule
    \end{tabular}
    \caption{A summary of the crucial colour image based toy datasets. Each dataset is given a code and a description and has its testing purpose outlined. The detection column contains the positive fraction detection; the control column shows the positive fraction for the control group and the uncoloured column shows the detection for the same images coloured monochrome.}
    \label{table:3d_cnn_datacodes}
\end{table*}

\begin{figure}
    \centering
    \includegraphics[width=0.9\linewidth]{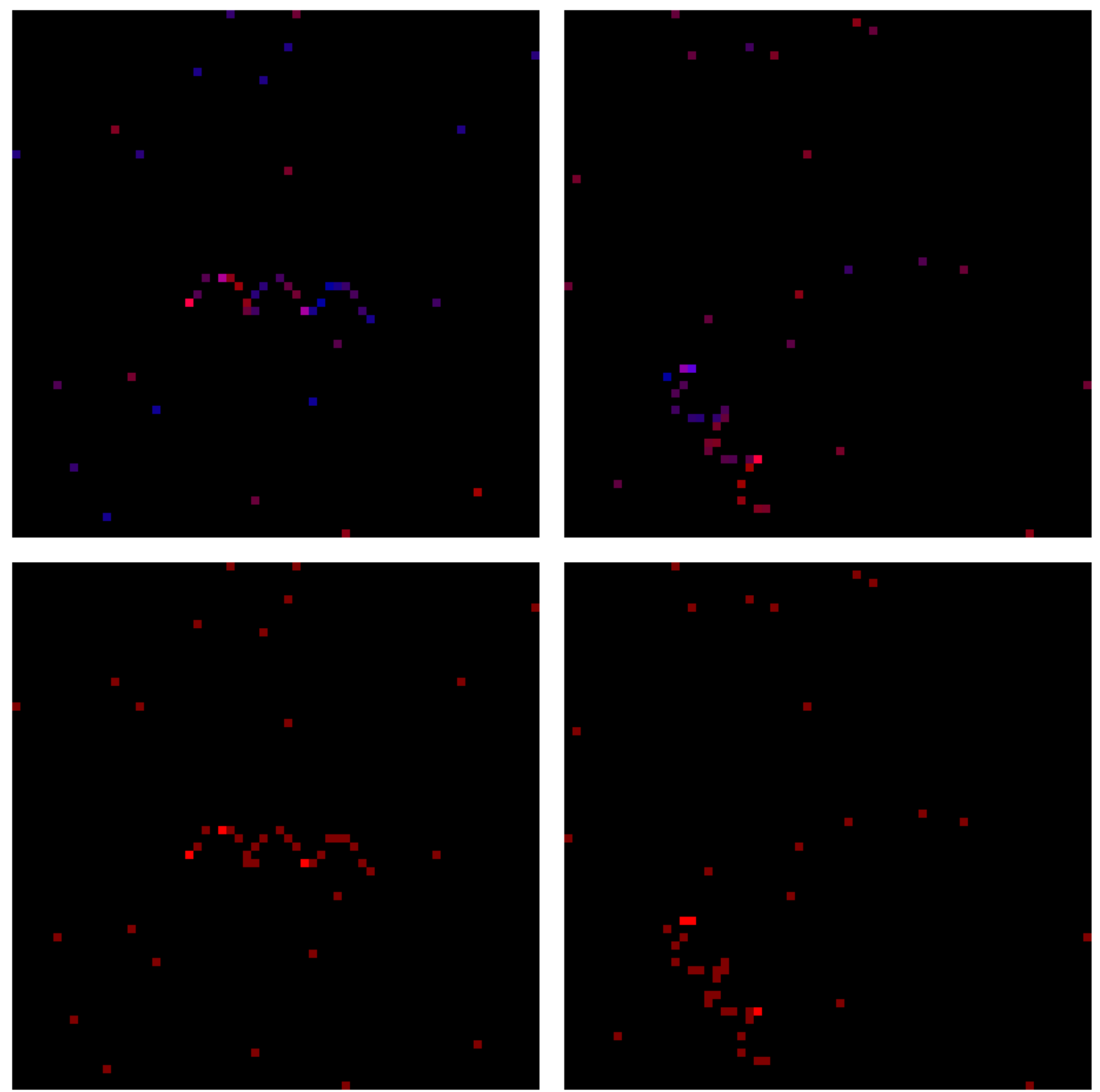}
    \caption{Sample images from a 3d dataset containing left-handed helices and a random field of points. \textbf{Top:} Images where points are coloured by their distance into the page. The 3d information creates a parity violating dataset. \textbf{Bottom:} All points are the same colour, the loss of information creates a dataset which is not parity violating.}
    \label{fig:helix_demonstration}
\end{figure}

To effectively test the functionality of this method, various datasets were constructed in three-dimensional space. The key results follow in \cref{table:3d_cnn_datacodes} and  a full list of the important test can be found in \cref{appendix: test_results.}. Because we are using the same net as for the 2d convolutions, we can directly compare the colouration scheme outlined in \cref{sec:3d_CNN} with inputs where all galaxies are monochrome points, as show in \cref{fig:helix_demonstration}.  
This shows significant improvement in detection, and results in detections that would not otherwise be made. These tests illustrate the success of the introduction of colour channels as a method for including the third dimensional information.

% \begin{figure}
%     \centering
%     \includegraphics[width=0.9\linewidth]{Illustration images/image_without_background.png}
%     \caption{A demonstration of creating pixelised colour images. In each row the distance coordinate increases from 0 to 1.0; this changes the colour from blue to red. The top three rows have weights 1.0, 0.5 and 0.25 respectively; this changes the intensity. In the fourth row each pixel contains two `galaxies', each with weight 0.25. The fifth row contains the same points as the third row, but each occupied pixel also contains a second `galaxy' further away than the first; this creates pixels with a redder tinge.}
%     \label{fig:pixel_generation}
% \end{figure}

\subsection{2D Angle based Method}\label{sec:angle}

To further the investigation into three dimensions, and facilitate testing a corollary to the 4PCF a new neural net was constructed with the same guiding principles as the CNN. Since the inputs were numeric, two options were considered: 1D convolutions or linear layers. We performed the tests on the toy data using both separately. We find that both perform equally well, but the linear layers are marginally faster, so these were chosen to be used on the real data.\\
We begin by considering the two-dimensional case. In 2d the lowest order polyhedron subject to parity violation is the triangle. A triangle in 2d space, is defined by six degrees of freedom (d.o.f.) - three two-dimensional coordinates - to which the constraints of rotational and scale invariance are added, these reduce the six d.o.f.\ down to two. Two element inputs are then sufficient, and a convenient choice is the internal angles of the triangle. To make the angles subject to parity violation, they require ordering in a way that would change under the action of a parity operator. The method used is outlined as follows for groups of 3 points:
\begin{enumerate} 
    \item Find the mean position of the group and shift it to be the origin
    \item Compute the distances of the points to the origin use this to order the points with the closest as the first
    \item Reorder the second and third in an anti-clockwise direction from the first
    \item Compute the angles between the points in an anti-clockwise fashion
    \item Define $\hat{P}$ to swap the first and third angles
    \item Use the first two angles in every group of 3 as the net inputs
\end{enumerate}

\begin{figure}
    \centering
    \includegraphics[width=\linewidth]{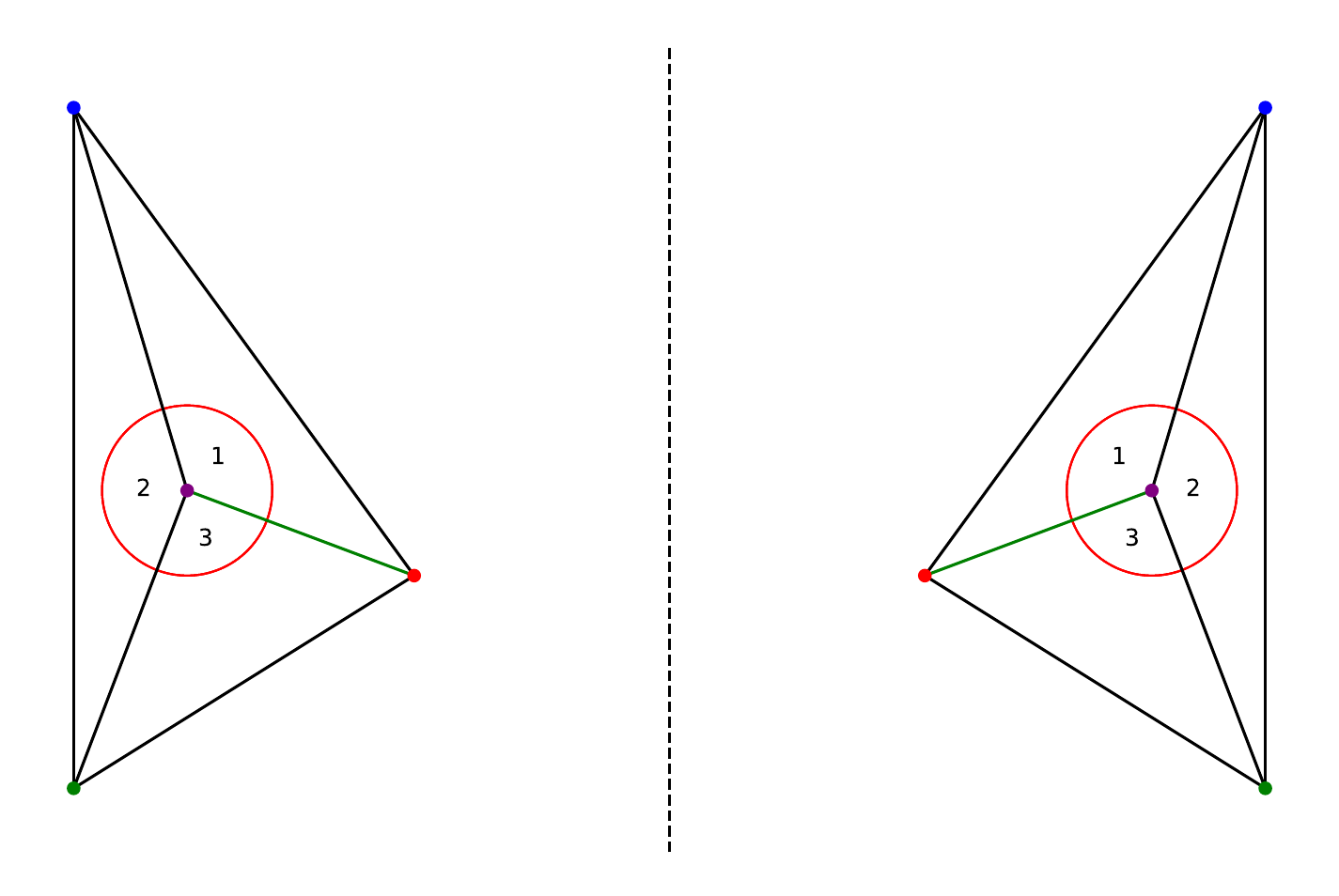}
    \caption{A demonstration of the angle method in 2d. The line connecting the closest vertex to the centroid is shown in green. Working anticlockwise from this line, the initial angles are ordered [1,2,3], under a parity transformation this changes to [3,2,1].}
    \label{fig:angles_method}
\end{figure}

The principles of this method are displayed in \cref{fig:angles_method}, which illustrates why this is the appropriate parity operator.
In fact this method is generally extensible to any $n$-gon in 2d, with the first $n-1$ angles being the input and the parity operator inverting the order of the angles as shown in \cref{eq:2d_n_angle}. Conveniently for the function of the neural net, the nth angle can be calculated from the others, as given in \cref{eq:nth_angle}.

\begin{equation}
    \hat{P}[\theta_1, \theta_2, ..., \theta_{n-1}] = [\theta_n, \theta_{n-1}, ..., \theta_2] 
    \label{eq:2d_n_angle}
\end{equation}

\begin{equation}
    \theta_n = 2\pi - \sum_{i=1}^{n-1}\theta_i
    \label{eq:nth_angle}
\end{equation}

\subsubsection{Testing method with toy data}

\begin{table*}
    \begin{tabular}{  p{2.35cm}  p{5.5cm}  p{4.9cm} P{2.0cm} P{2.0cm}}
        \toprule
\textbf{Data code}      
& \textbf{Description}   
& \textbf{Purpose} 
& \textbf{Detection}
& \textbf{Control}\\\midrule
R2d
& Random Generated 2d point Data      
& Check parity violation is not detected where there is none
& $50.41\pm0.10\%$ 
& $50.10\pm0.09\%$ \\\hline
PV-0.1-R2d    
& R2d with two extra points generated near each R2d point, these are generated in a way that all groups of three are of the same chirality. Groups of three are separated on the scale of one tenth the average inter-point distance
& Check detection of parity violating inputs
& $90.77\pm0.06\%$ 
& $49.74\pm0.10\%$ \\\hline
S-PV-0.1-R2d
&PV-0.1-R2d, but after generating parity violating groups all points shuffled together and new groups drawn at random
&Can parity violations be detected when they are visibly present, and groups sampled randomly from the dataset
& $49.70\pm0.10\%$ 
& $49.68\pm0.10\%$ \\\hline
C-PV-0.1-R2d
&S-PV-0.1-R2d, with a distance constraint when generating groups from the shuffled data   
& Can parity violations become detectable by constraining the scale
& $65.10\pm0.04\%$ 
& $49.85\pm0.04\%$ \\\hline
\end{tabular}
    \caption{2D toy datasets. Each dataset is given a code and a description and has its testing purpose outlined. The detection column contains the positive fraction detection, and the control column shows the positive fraction for the control group for each test.}
    \label{table:2d_ang_datacodes_summary}
\end{table*}

\begin{figure}
    \centering
    \includegraphics[width=\linewidth]{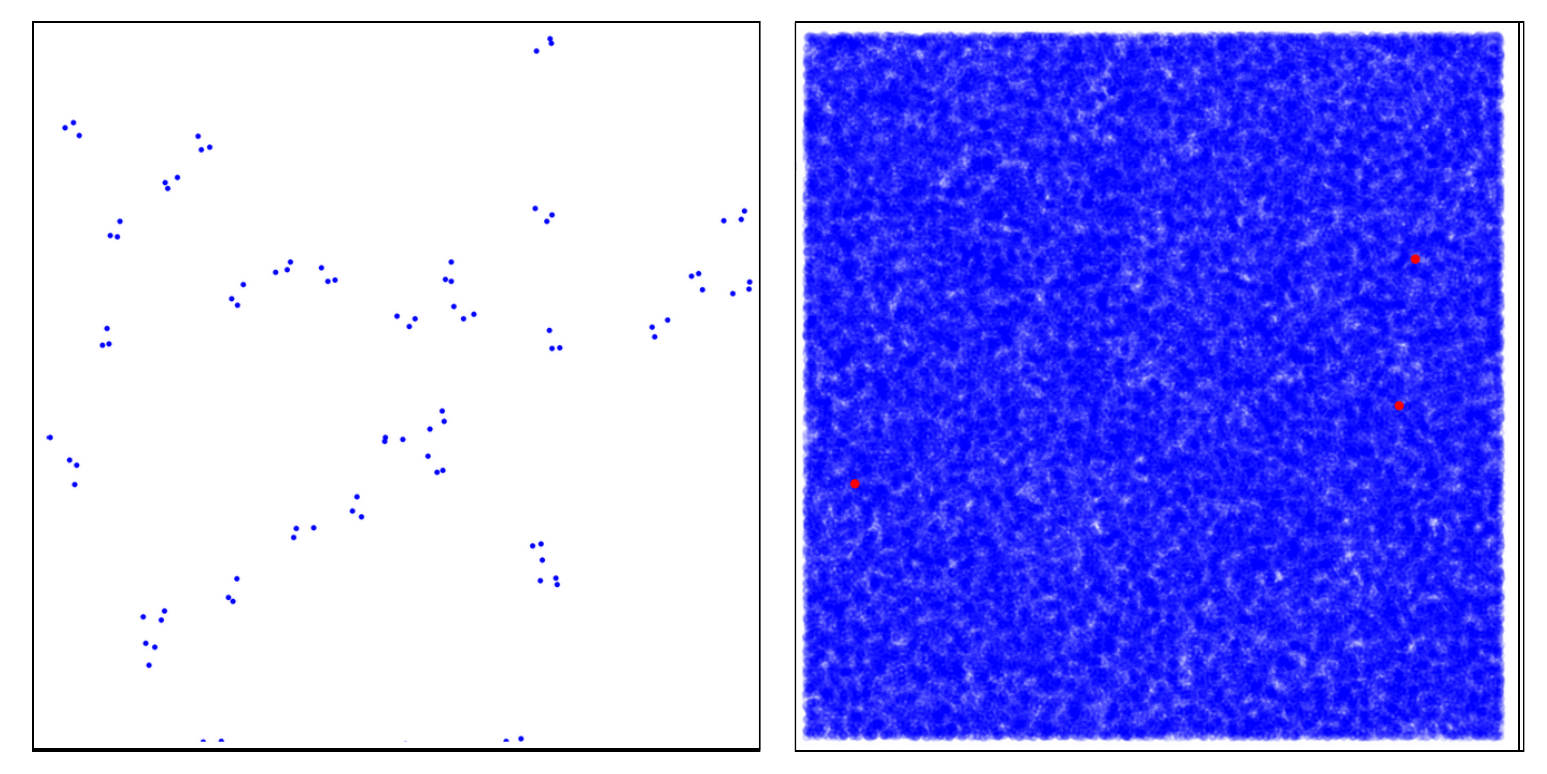}
    \caption{An illustration of the failure of the 2 dimensional angles method to find parity violations across arbitrary scales. \textbf{Left:} A localised region of the field demonstrating the observable parity violation in the data. \textbf{Right:} The full field of points, with a randomly sampled group of 3 three shown in red. This demonstrates the impossibility of finding the parity violation when sampling across any scale: the chance of picking three points that exist in a parity violating way is negligible.}
    \label{fig:2d_angle_failure}
\end{figure}

\begin{figure}
    \centering
    \includegraphics[width=\linewidth]{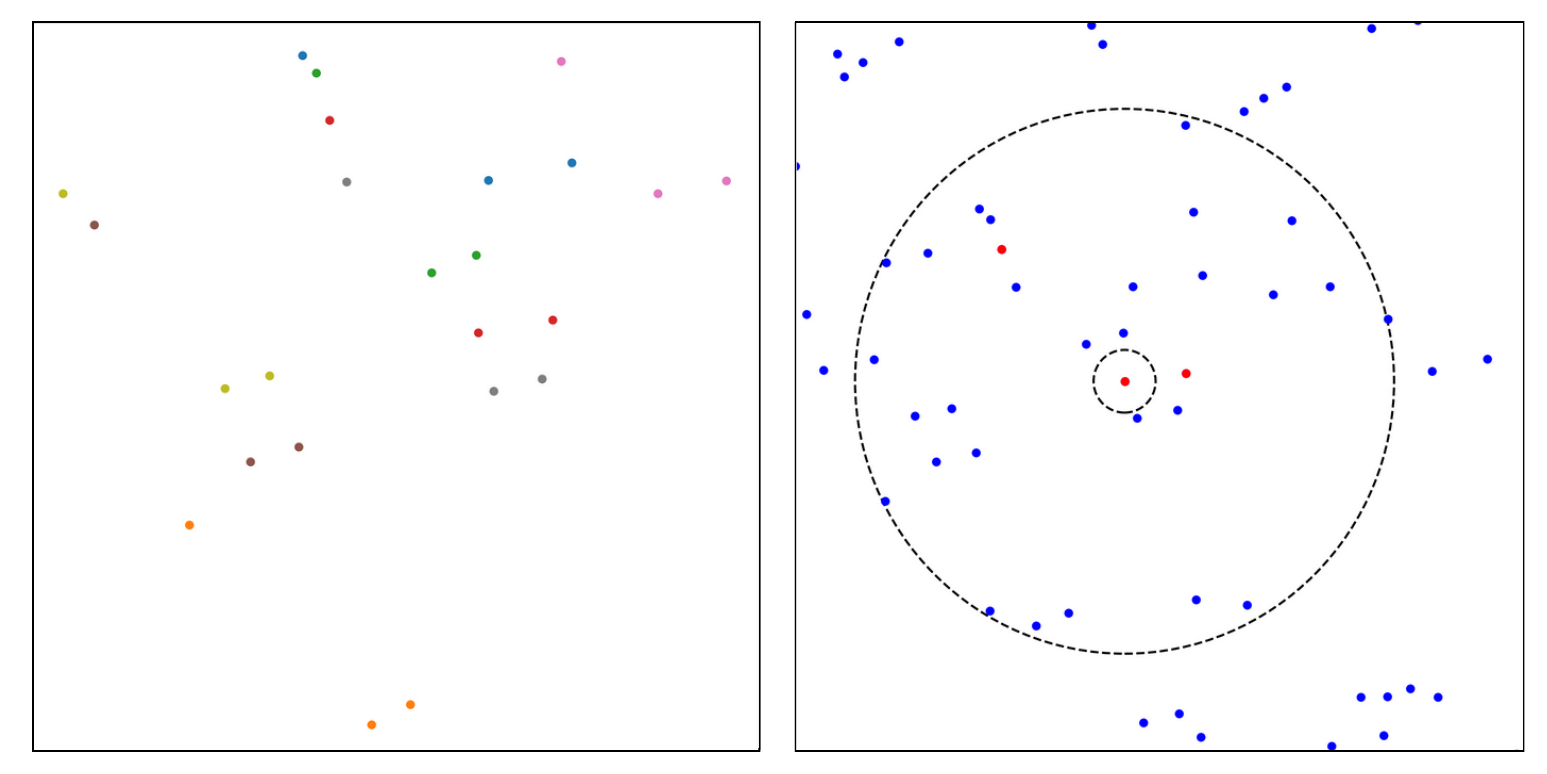}
    \caption{By constraining the scale on which to construct a group of points, the net has a significantly improved likelihood of detection. \textbf{Left:} A localised region of the field of points. The field is chiral by construction, and each group of three has been coloured to show the asymmetry. \textbf{Right:} Adding a constraint on the scale over which groups can be constructed significantly increases the chances of detection.}
    \label{fig:2d_angle_constraint}
\end{figure}

As with the CNN, our first step was to test the efficacy and sensitivity of this method. This began with testing a random field and then working through a number of toy data scenarios, the critical ones are outlined in \cref{table:2d_ang_datacodes_summary}, and the full list can be found in \cref{appendix: test_results.}. These results crucially demonstrate that this method does not detect parity violation if not present, and can detect parity violation if it is present in at least 1\% of the input data. The main drawback of this method is that for it to be successful, the input data must be sampled such that it faithfully represents the parity of the system. Conceptualise this by considering a random uniform field of two-dimensional data points. For every point in the field generate two more points, positioned relative to the first so that they form a chiral triangle, and enforce the average separation of the points in each group of three to be one tenth of the average inter-point distance. This creates a visually parity-violating dataset, as displayed in \cref{fig:2d_angle_failure}. If the inputs to the net are calculated from these groups of three, there is a strong parity detection, as expected. However, the real data will not have the benefit of being pre-selected into parity violating groups, and therefore to produce a fair test it is necessary to generate the field as before, and then sample groups of three randomly from the total field. In this scenario, it becomes almost impossible to find the parity violation, because each point in a group is being sampled from a 500,000-point field, so the chances of selecting a group that represents the chirality is negligible. This is shown in \cref{table:2d_ang_datacodes_summary} with datasets \textbf{PV-0.1-R2d} and \textbf{S-PV-0.1-R2d}.
This issue can be mitigated by requiring samples to be drawn on a given scale. With this additional requirement, detection of parity violation becomes significant, as long as the scale constraint matches the scale on which the parity violation exists. This idea is demonstrated in \cref{fig:2d_angle_constraint} and shown in \cref{table:2d_ang_datacodes_summary} with dataset \textbf{C-PV-0.1-R2d}. Testing over different constraint scales, the data can be investigated in an analogous way to power spectrum analyses.

\subsubsection{Summary of tests}
After adding the constraint on the scale of the groups, a detection is made with a high level of significance if the constraint appropriately selects groups of data with at least $\sim1.0\%$ containing parity violation.
These tests were also extended to groups of 4 and 5 points, the full results of which can be found in \cref{appendix:2d_toy_test}

\subsection{3D Angle based Method}

\begin{figure}
    \centering
    \includegraphics[width=\linewidth]{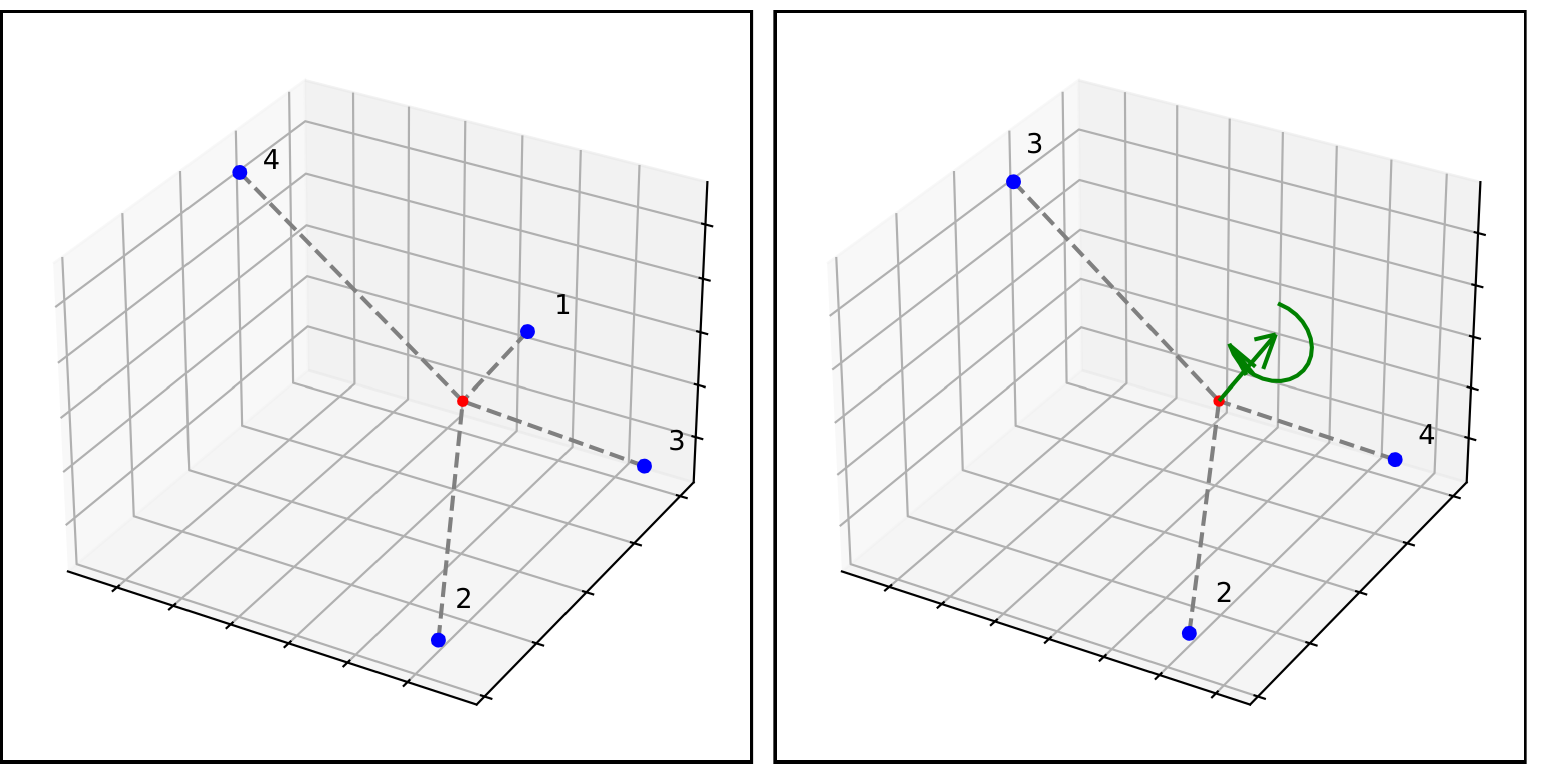}
    \caption{An illustration of the ordering of angles in 3 dimensions. The images show a group of four points with their mean position in red. \textbf{Left:} The points are initially ordered by distance from the mean. \textbf{Right:} The ordering is finalised by starting from the second point and working in a right-handed manner about the vector describing the position of the first point. In this case points three and four are reordered. }
    \label{fig:3D_angle_demo}
\end{figure}

Now that the success of an angle based method in 2d has been demonstrated, this motivates exploration of a 3d equivalent. To detect parity violation, groups of four points are the smallest scale on which this method will work, and so this can act as a direct comparison to the 4PCF. As in the two-dimensional case, the critical part of this method is defining an ordering of the angles and understanding how this ordering changes under a parity operator. There are a number of ways to define an ordering, the process used for groups of four points is outlined as follows, and displayed in \cref{fig:3D_angle_demo}:

\begin{enumerate}
    \item Compute the mean position of the group and shift it to be the origin
    \item Calculate the distances of the points to the origin, and order the points by distance with the closest point as the first
    \item Use the first position as an axis, and starting from the second-closest point, work around the axis in a right-handed manner to determine the order of points 3 and 4
    \item Compute the internal angles and order them as $[\theta_{12}, \theta_{13}, \theta_{14}, \theta_{23}, \theta_{34}, \theta_{24}]$
\end{enumerate}

This method fixes the positions of the first two points because that only depends on their distance to the mean. The ordering of the third and fourth points is dependent on the chirality of the group of points. This suggests an appropriate parity operator would swap positions three and four, which would reorder the angles as shown in \cref{eq:3d_angle_P}.

\begin{equation}\label{eq:3d_angle_P}
    \hat{P}[\theta_{12}, \theta_{13}, \theta_{14}, \theta_{23}, \theta_{34}, \theta_{24}] = [\theta_{12}, \theta_{14}, \theta_{13}, \theta_{24}, \theta_{34}, \theta_{23}]
\end{equation}

To ensure this is always the appropriate parity operator, we generated 100,000 tetrahedra, and determined their orders by the process above, giving [1, 2, 3, 4]. Parity operators in three-dimensional space were defined: $\hat{P}_x, \hat{P}_y, \hat{P}_z$, flipping in the $yz,\ zx\ \text{and}\ xy$ planes respectively. Combinations of these parity operators were applied to each tetrahedral group. The four combinations that do not change the parity are $[\mathbb{1},\ \hat{P}_x\hat{P}_y,\ \hat{P}_y\hat{P}_z,\ \hat{P}_z\hat{P}_x]$ and these all leave the order as [1, 2, 3, 4]. The four parity flipping combinations are $[\hat{P}_x,\ \hat{P}_y,\ \hat{P}_z,\ \hat{P}_x\hat{P}_y\hat{P}_z]$ and all were found to change the order of the points to [1, 2, 4, 3]. This confirmed that the parity operator given in \cref{eq:3d_angle_P} is the correct operator in this angular description.\\
A group of 4 points in three-dimensional space has 12 degrees of freedom, - 4, 3d coordinates - and after accounting for rotational and scaling invariance, this reduces to 5 d.o.f. Since a tetrahedron has 6 internal angles ($\Comb{6}{2}$), the inputs are chosen to be the first 5: $[\theta_{12}, \theta_{13}, \theta_{14}, \theta_{23}, \theta_{34}]$. 
In the 2d case the final angle, which is needed for the parity transformation, can be calculated from the others, as shown in \cref{eq:nth_angle}. In the 3d case it is not so simple. Trying to use the first five angles gives four choices for the sixth. To overcome this issue, we created a custom data class, so that every data entry has all six internal angles, the first five of which provide the basic inputs to the net, and the sixth, which can be accessed to generate the parity transformed inputs.

\subsubsection{Testing method with toy data}\label{sec:test_3dang}
Once again, the significant work in creating the net was establishing its accuracy and sensitivity. Numerous tests were carried out, with the results listed in \cref{table:3d_ang_datacodes_summary}.

\begin{table*}
    \begin{tabular}{  p{2.35cm}  p{5.5cm}  p{4.9cm} P{2.0cm} P{2.0cm}}
        \toprule
\textbf{Data code}      
& \textbf{Description}   
& \textbf{Purpose} 
& \textbf{Detection}
& \textbf{Control}\\\midrule
R3
& Random Generated 3d point Data      
& Check parity violation is not detected where there is none
& $49.90\pm0.11\%$ 
& $50.11\pm0.10\%$ \\\hline
PV-0.1-R3    
& R3 with three extra points generated near each R3 point, these are generated in a way that all groups of four are of the same chirality. Groups of four are separated on the scale of one tenth the average inter point distance
& Check detection of parity violating inputs
& $87.59\pm0.11\%$ 
& $49.69\pm0.07\%$ \\\hline
S-PV-0.1-R3
&PV-0.1-R3, but after generating parity violating groups all points shuffled together and new groups drawn at random
&Can parity violations be detected when they are visibly present, and groups sampled randomly from the dataset
& $50.37\pm0.08\%$ 
& $50.15\pm0.08\%$ \\\hline
C-PV-0.1-R3
&S-PV-0.1-R3, with a distance constraint when generating groups from the shuffled date   
& Can parity violations become detectable by constraining the scale
& $58.73\pm0.08\%$ 
& $49.87\pm0.08\%$ \\\hline
C-PV-1-R3  
&PV-0.1-R3, but scale of parity violations now same scale as average inter-point distance. No longer visibly parity violating. Groups drawn with scale constraint
& Check parity violations detectable on more realistic scales with scale constraint
& $54.37\pm0.05\%$ 
& $49.87\pm0.06\%$ \\\bottomrule
    \end{tabular}
    \caption{3D toy datasets. Each dataset is given a code and a description and has its testing purpose outlined. The detection column contains the positive fraction detection, and the control column shows the positive fraction for the control group for each test.}
    \label{table:3d_ang_datacodes_summary}
\end{table*}

The main conclusions of the tests were that the method does detect significant parity violations, as long as at least 1.2\% of the inputs are net parity violating. The main drawback of the method is the difficulty of ensuring the groups of four points are sampled such that if there is parity violation, it is appropriately represented. The dataset \textbf{S-PV-0.1-R3} shows that some constraints are needed, and as in the 2d case, it was shown (\textbf{C-PV-0.1-R3}) that constraining the scale on which groups are generated is sufficient to capture the parity violation, as long as the scales are aligned. Group \textbf{C-PV-1-R3} shows that parity violations are detected where the human eye would be unable to, a powerful demonstration of this method.

\subsection{Vector input method}

To finalise the search for parity violations at the 4-point level, we construct a vector method. For this net the inputs are components of the vectors describing a group of points in their mean at origin frame. To account for the scaling invariance, all inputs are normalised to make the longest vector in the group have a length of unity. The rotational invariance of the setup is accounted for in the construction of $f(x)$:

\begin{multline}
    f(x) = \sum_{\hat{R}_i}\ \left[g(\hat{R}_ix)\ -\ \sum_{i}g(\hat{R}_i\hat{P}_ix)\ + \right. \\ \ \left. \frac{1}{2}\sum_{i \neq j}g(\hat{R}_i\hat{P}_i\hat{P}_jx)\ -\ g(\hat{R}_i\hat{P}_x\hat{P}_y\hat{P}_zx)\right]  
    \label{eq:strange}
\end{multline}
where $\hat{R}_i \in [\hat{R}_{0}, \hat{R}_{90}, \hat{R}_{180}, \hat{R}_{270}]$. This ensures that every orientation of each handedness is equally represented.

\subsubsection{Testing method with toy data}
Because this also considers 3d distributions, we used the same tests as those in \cref{sec:test_3dang} on the 3d angle method. The results are listed in \cref{table:3d_vec_datacodes_summary} and show the efficacy and suitability of this method. As with the previous geometric methods, the only way to detect parity violations is to issue a constraint on the scale of the testing groups. It is important to note that the deviations from 50\% are much smaller in this case, because of the sampling over 3d, thus significance is shown using $\sigma$-levels in \cref{sec:results}. 

\begin{table}
\centering
    \begin{tabular}{P{2.5cm} P{2.2cm} P{2.2cm}}
        \toprule
\textbf{Data code}       
& \textbf{Detection}
& \textbf{Control}\\\midrule
R3
& $50.11\pm0.07\%$ 
& $50.21\pm0.06\%$ \\\hline
PV-0.1-R3    
& $99.83\pm0.02\%$ 
& $50.3\pm0.31\%$ \\\hline
S-PV-0.1-R3
& $50.16\pm0.07\%$ 
& $49.84\pm0.07\%$ \\\hline
C-PV-0.1-R3
& $52.44\pm0.10\%$ 
& $50.11\pm0.11\%$ \\\hline
C-PV-1-R3  
& $52.68\pm0.07\%$ 
& $49.91\pm0.07\%$  \\\bottomrule
    \end{tabular}
    \caption{Results for the vector input method on the 3d toy datasets, as defined in \cref{table:3d_ang_datacodes_full} The detection column contains the positive fraction detection, and the control column shows the positive fraction for the control group for each test. }
    \label{table:3d_vec_datacodes_summary}
\end{table}

\subsection{Summary of Sensitivities}
After extensively testing each method, the net fraction of the inputs required to contain parity violating objects in order to make a detection significant from non-zero was determined. The level for each method is outlined in \cref{tab:frac_detect}. For a null detection, these are the levels above which we can conclude that parity violations are not present.

\begin{table}
\centering
    \begin{tabular}{c c}
        \toprule
\textbf{Method}     
& \textbf{Fraction to detect}\\\midrule
2D Images
& $<0.1\%$ \\\hline
3D Images
& $\sim0.5\%$ \\\hline
2D Angles
& $\sim1.0\%$\\\hline
3D Angles
& $\sim1.0\%$ \\\hline
3D Vectors
& $\sim2.0\%$\\\hline

    \end{tabular}
    \caption{Summary of the required net fraction of the inputs containing a ``parity violating'' object for the network to return a fraction significantly different from random input data.}
    \label{tab:frac_detect}
\end{table}

\section{Generating inputs and Results}\label{sec:results}

\subsection{Curation for 2d CNN}
To generate images from the BOSS catalogue the method outlined in \cref{sec:2d_conv} was used. 12,000 images were sampled, split into 2,400 testing and 9,600 training images, each covering a 0.5\degree x 0.5\degree\ patch of the sky. This size was chosen to create a large enough image dataset that suitably covered the full field of galaxy data. Smaller images also offset issues with approximating a spherical surface onto a flat projection and reduce overlap of images onto survey artifact regions. A full list of tests on different sizes can be found in \cref{appendix:image_sizes}.
\begin{figure}
    \centering
    \includegraphics[width = \linewidth]{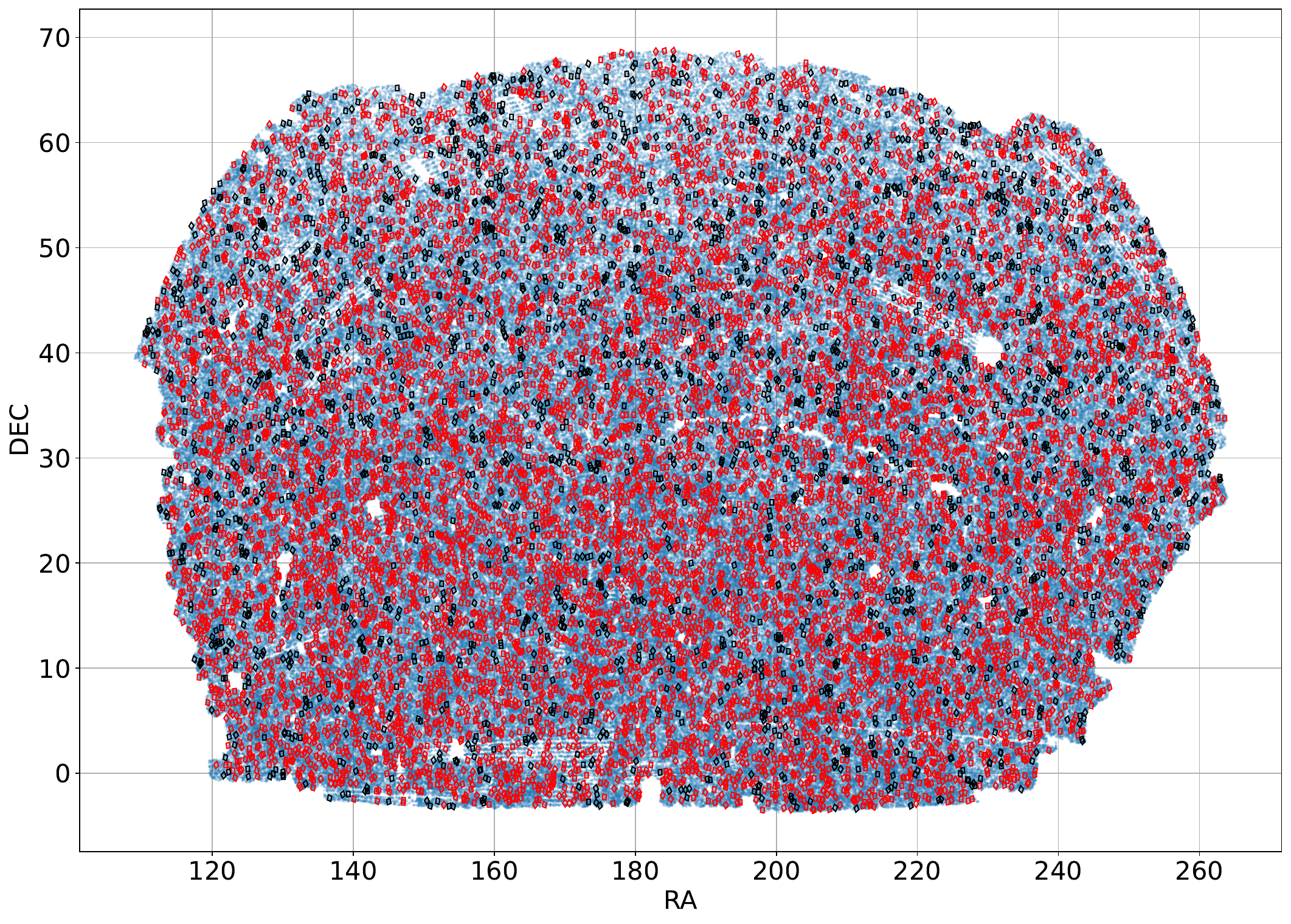}
    \caption{A sample of images taken from CMASS. The black squares show testing images, and the red squares show training images.}
    \label{fig:CNN_demo_images}
\end{figure}

\subsection{Curation for 2d CNN with colour}
To generate the 3d images, the method outlined in \cref{sec:3d_CNN} was used. Each voxel of the space had a side length of 0.5\degree and a redshift depth of 0.33. Since there is no approximation of the 3d distribution, there is no constraint on the upper size of the images, and so varying image sizes were tested.

\subsection{Curation for geometric nets}
For the 2d angle method, groups were constructed from the 2d projection of the data, with points in a group constrained to be within 1\degree\ of each other. Groups were constructed on the 3, 4 and 5 point levels. In the 3d case, for both the angle and vector input methods, groups were constructed to contain 4 points, allowing for direct comparison to the 4PCF method. These groups were drawn from the CMASS catalogue, with the constraint that the points must be separated by a distance in the range of $\ 20h^{-1}$Mpc to $160h^{-1}$Mpc. In all cases datasets of 500,000 data groups were constructed; 400,000 for training and 100,000 for testing.

\subsection{Results}
The average level of detection from bootstrapping over the testing sets and their controls for each method are given in \cref{tab:method_results}. We also include results from running over the random catalogue provided by BOSS. For every method, no significant deviation from the control or random data is detected. This is confirmed in the $\sigma$ levels displayed in \cref{tab:t_test_results}. We do not find evidence of parity violation.

\begin{table}
\centering
    \begin{tabular}{P{1.6cm}  P{1.85cm} P{1.85cm} P{1.85cm}}
        \toprule
\textbf{Method}      
& \textbf{Detection}   
& \textbf{Control} 
& \textbf{Random}\\\midrule
2D Images
& $50.4\pm0.3\%$
& $50.1\pm0.3\%$ 
& $50.8\pm0.3\%$\\\hline
Colour Images
& $49.7\pm0.3\%$
& $49.4\pm0.3\%$
& $49.3\pm0.3\%$ \\\hline
2D Angles, n=3
& $50.15\pm0.09\%$
& $50.02\pm0.09\%$
& $50.35\pm0.10\%$\\\hline
2D Angles, n=4
& $50.05\pm0.06\%$
& $49.85\pm0.06\%$
& $50.20\pm0.06\%$\\\hline
2D Angles, n=5
& $50.30\pm0.06\%$
& $50.27\pm0.06\%$
& $49.65\pm0.06\%$\\\hline
3D Angles
& $49.91\pm0.06\%$
& $50.09\pm0.05\%$ 
& $49.84\pm0.05\%$\\\hline
3D Vectors
& $50.08\pm0.03\%$
& $50.09\pm0.03\%$
& $50.13\pm0.03\%$\\\hline

    \end{tabular}
    \caption{The average deviation from parity even of the CMASS NGC catalogue for the different methods. Results for the control groups and random catalogue are also shown.}
    \label{tab:method_results}
\end{table}

\begin{table}
\centering
    \begin{tabular}{c c c}
        \toprule
\textbf{Method}      
& \textbf{CMASS-Control}   
& \textbf{CMASS-Random}\\\midrule
2D Images
& 1.00
&1.33 \\\hline
Colour Images
&1.00
&1.33 \\\hline
2D Angles, n=3
& 1.44
&  2.00\\\hline
2D Angles, n=4
& 1.67
&  2.50\\\hline
2D Angles, n=5
& 0.500
&  0.833 \\\hline
3D Angles
&  0.00
&  1.00\\\hline
3D Vectors
& 0.333
&  1.67\\\hline

    \end{tabular}
    \caption{$\sigma$-level significances calculated for  both (i) the difference between the CMASS data and the CMASS controls, and (ii) the difference between the CMASS data and the BOSS random catalogue data. No significant parity violation is found.}
    \label{tab:t_test_results}
\end{table}

\section{Discussion}

The results suggest there is no parity violation to be found in the LSS represented in the NGC of the CMASS catalogue. For the 2d projections, parity violation is not expected, because compression of a three-dimensional dataset down to two dimensions removes information about the three-dimensional distribution.

For the three-dimensional cases, the results obtained are at odds with the work of \citet{Philcox_2022} and \citet{Hou_2023}. Recent analyses which account for the 8PCF bias term, have also disputed the detection of parity violation~\citep{krolewski2024evidenceparityviolationboss}. During a recent talk at the Royal Society~\citep{RS_A}, Philcox announced results that cast doubt on the parity violation detection~\citep{philcox2021detection, Philcox_2022}, which will be released in an upcoming issue of \citet{PT_RSA}.

\begin{figure}[t!]
    \centering
    \includegraphics[width=\linewidth]{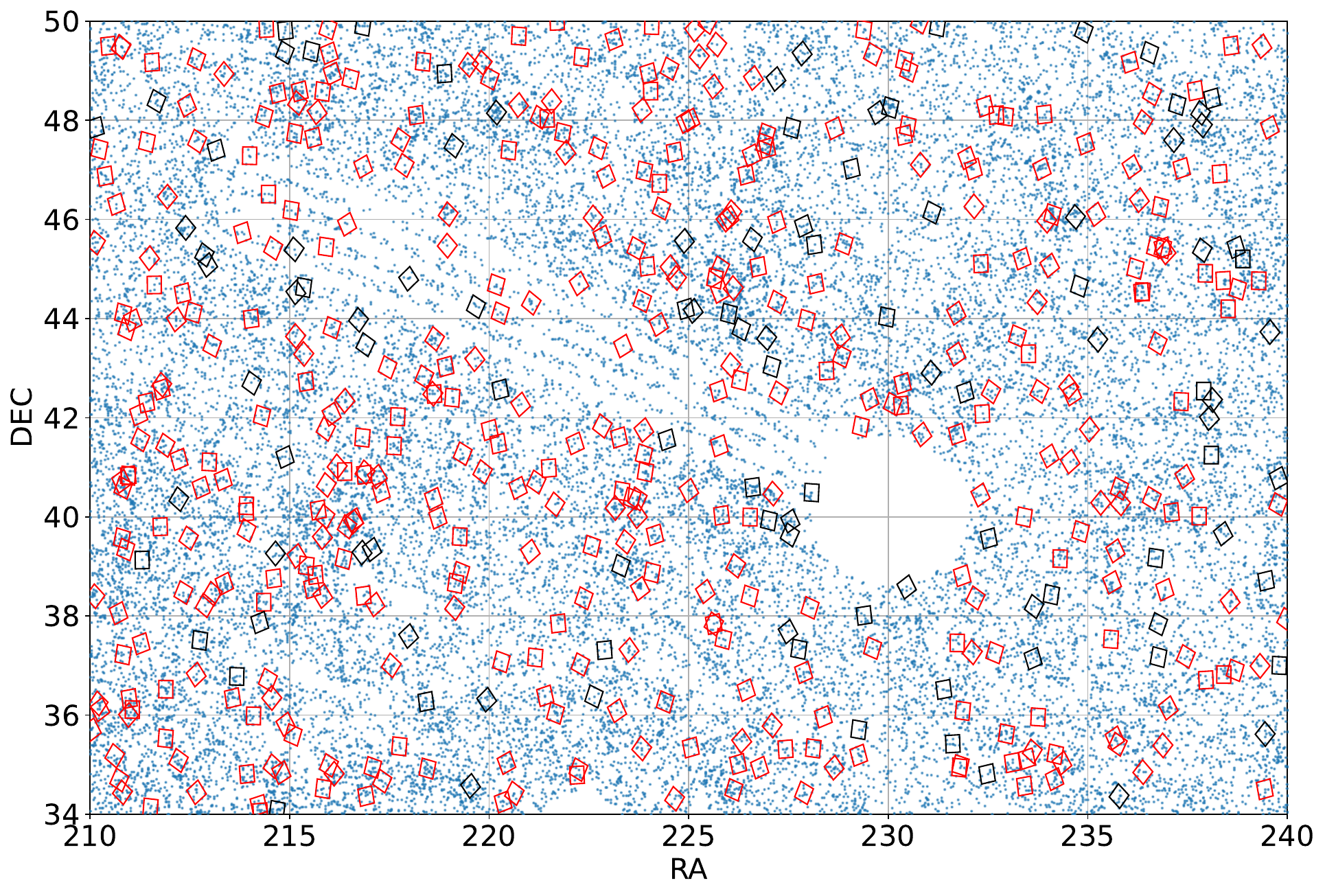}
    \caption{The set of images generated from points near the large artifact between (210, 35) and (240, 50). Few images are affected by the presence of the artifact, and those which are affected have small overlaps into the affected region.}
    \label{fig:Artifact_distribution}
\end{figure}

\begin{figure}[t!]
    \centering
    \includegraphics[width=\linewidth]{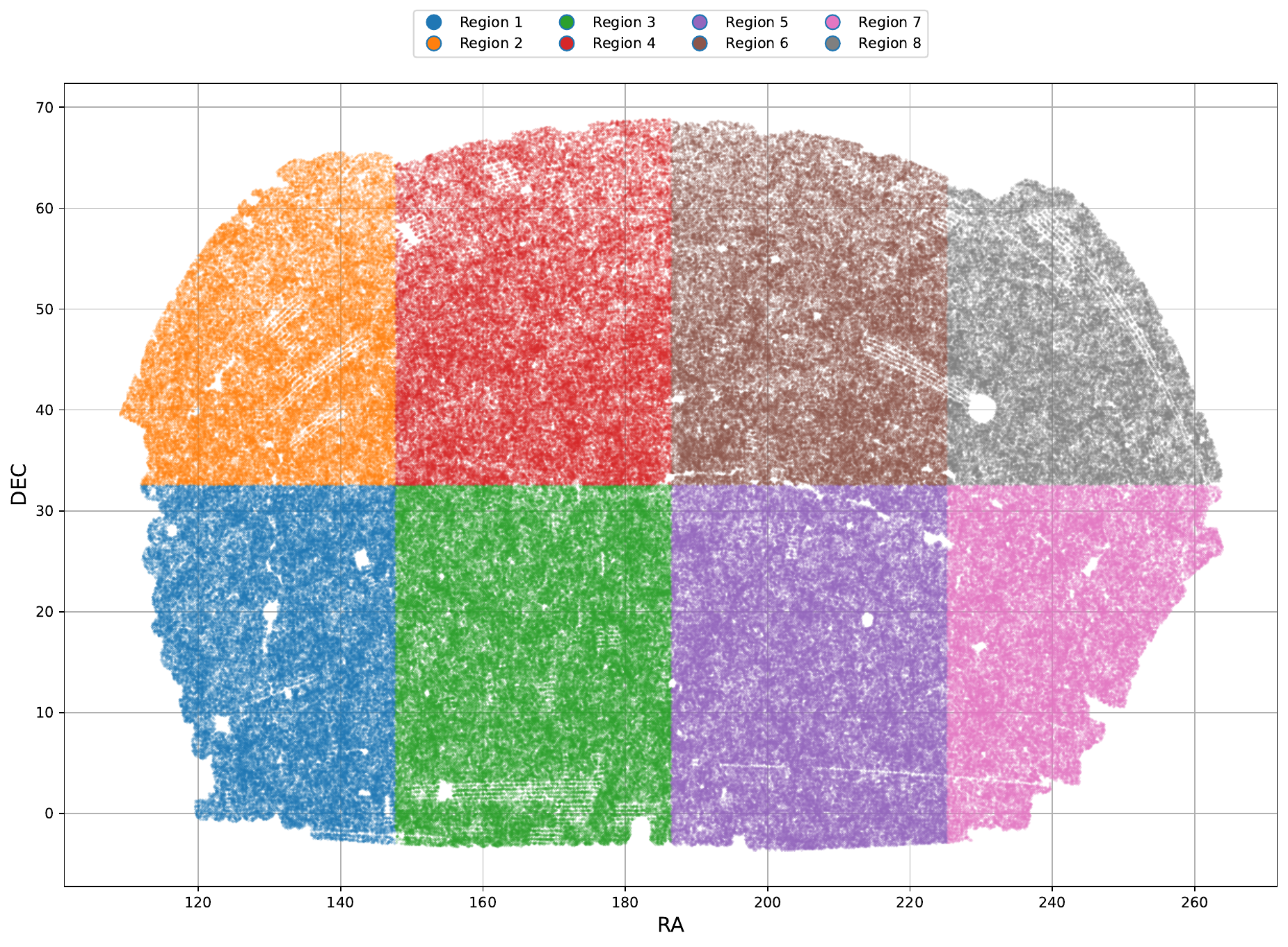}
    \caption{The CMASS catalogue split into 8 regions, each of which was investigated separately. Each region showed the same level of parity violation, as shown in \cref{appendix:binned_sky}. This further suggests that the artifacts are not skewing the results.}
    \label{fig:Split_cmass}
\end{figure}

An important consideration is the `artifacts' of the dataset. The 2d projection in \cref{fig:cmass_NGC} displays notable sample artifacts, which could interfere with the analyses. In all methods, the weights applied \cref{sec:Data} act to account for sampling issues, reducing the impact. For the CNN, images were constructed to further offset the impact: sampling a large enough set of small enough patches minimised overlap with artifact regions, displayed in \cref{fig:Artifact_distribution}.  That the overall impact is negligible is perhaps best displayed by the null results in all cases. Given that no parity violation is detected, if the artifacts were interfering, it would have to be in a way that perfectly cancels with some underlying parity violation of the dataset. Since no detection is made in either 2d or 3d, the probability of this being the case is negligible. To further the rigorousity of these claims, the analyses were performed over the random catalogue of the BOSS dataset. This catalogue is constructed to match the sampling geometry of the BOSS data, but is $\sim$50 times the size. To resolve this size mismatch, the random data was split into 50 chunks, 10 of which were used for the analysis. The results given in \cref{tab:t_test_results} show that sampling geometry indeed introduces no parity violation.
To make a final confirmation, the sky was split into 8 bins by position, displayed in \cref{fig:Split_cmass}, and inspired by \citet{Hou_2023}. Treating each of these separately, no significant difference was found between regions (\cref{appendix:binned_sky}). Since every region is differently affected by artifacts, this suggests they have no noticeable impact on results.

A final important consideration is to note that the errors quoted for each dataset represent the model variance arising from our network and the resampling process. Cosmic variance from the single realisation of the Universe will contribute an extra source of uncertainty to the BOSS catalogue, such that even in the limit of infinite samples of the data, some underlying non-zero baseline variance would remain.

Given the sensitivity of the methods, and the issues with searching on different scales, the CNN seems to be the most suitable method for these analyses. Its main advantage is that each image contains the full distribution for a region of space, making it the most general search method for parity violations across scales. In this work we proposed and demonstrated the effectivity of conveying three-dimensional information through colour creating a cCNN. We do not claim that this is the most suitable method, but the success of our test cases provides evidence to show the power of this method. To further this pursuit, convolutions could be extended into three dimensions; extensive further work would be required to ensure the model would correctly detect chirality. A possible implementation of this three-dimensional structure could be layers of two-dimensional images, stacked into three-dimensional voxel inputs, similar to the application in classification of medical images~\citep{med_3dCNN1, singh20203d_CNN}.

\section{Conclusion}

We used machine learning methods to search for parity violations in the Large-Scale Structure of the Universe. Five variations of the method were employed, two investigating the two-dimensional projection of the data, and three investigating the three-dimensional distribution. It was shown that all of these methods can detect chiral datasets down to significance fractions outlined in \cref{tab:frac_detect}. Over various samples from the BOSS CMASS catalogue, no parity violation was detected in any of these methods. As such, we cannot provide further support for the recent claims by \citet{philcox2021detection} and \citet{Hou_2023}. 
To consolidate the work outlined in this paper, the CMASS SGC and LOWZ catalogues should be explored. Uncertainties in the fiducial cosmology used (\cref{sec:Data}) would alter the spatial distribution in three dimensions, and so the implications of these uncertainties should be investigated. Additionally, it would be useful to directly compare these methods to the 4PCF, and see how each performs against mock parity violating datasets. Further work should also be made in creating a three-dimensional CNN architecture to better explore local regions for parity violations across all orders in three dimensions. All of these methods should be applied to upcoming DESI, EUCLID and Roman data releases.

\begin{acknowledgments}

SH thanks the Cavendish Laboratory for the Project opportunity. SH also thanks St John's College and the University of Cambridge for funding.
WH was supported by a Royal Society University Research Fellowship.

\end{acknowledgments}

\section{Data Availability}
The data that support the findings of this article are openly available \cite{boss_data, Hewson_Unsupervised_searches_for_2025}.

% The \nocite command causes all entries in a bibliography to be printed out
% whether or not they are actually referenced in the text. This is appropriate
% for the sample file to show the different styles of references, but authors
% most likely will not want to use it.
%\nocite{*}
\newpage
\bibliographystyle{unsrtnat}
\bibliography{apssamp}% Produces the bibliography via BibTeX.
\newpage
\appendix

\section{Image size and type tests}\label{appendix:image_sizes}

The work carried out by \citet{lester_gibbon2022} demonstrated that using different image types has no effect on the function of the net, JPEG and PNG images could both be used equally. We chose to use PNG images because space-saving was not an issue, and the higher quality would make the pngs easier to work with.

When investigating the size, the primary considerations were:
\begin{enumerate}
    \item Does each image contain enough points to be able to detect parity violation on real scales. We required $3 < \braket{N}$, since we need at least 3 points to create a chiral object in 2d. 
    \item Are images small enough to get a large sample of images, where training and testing images do not overlap, and each set suitably represents the data.
    \item Are images small enough to minimise effects of approximating 3d distribution to 2d plane.
\end{enumerate}

These considerations led to square side lengths in the range ${0.3 < l < 1.0}$. A few sizes were tested on real and random data, the results of which are given in \cref{tab:image_size_results}. No method had a specific advantage in testing,  so we chose to use 0.5\degree\ square images because whilst larger point clouds could be used, 22.3 points is a reasonable level to look for parity violations whilst minimising any effects of projecting onto 2d space. 
\begin{table}
\centering
    \begin{tabular}{c c c c}
        \toprule
\textbf{Side length}      
& \textbf{Avg. points}   
& \textbf{Detection}
& \textbf{Rand. Detection}\\\midrule
$l = 0.3$
& 8.0
& 50.8
& 51.2\\\hline
$l = 0.5$
& 22.3
& 50.1
& 50.6\\\hline
$l = 0.7$
& 43.7
& 51.1
&50.5\\\hline
$l = 1.0$
&89.2
& 50.1
& 49.3\\\hline
    \end{tabular}
    \caption{Test results on different size image patches. On each size, tests were run on the real CMASS data and randomly generated data.}
    \label{tab:image_size_results}
\end{table}

\section{Important test results}\label{appendix: test_results.}

The following tables list more fully the important results from the validation of the methods. \cref{table:3d_cnn_datacodes_full} gives the results from the cCNN; \cref{table:2d_ang_datacodes_full} the results from the 2d angle method; \cref{table:3d_ang_datacodes_full} results from the 3d angle method; and \cref{table:3d_vec_datacodes_full} the results from the 3d vector method.

\begin{table*}
    \begin{tabular}{ p{2.0cm}  p{5.1cm}  p{4.2cm} P{1.9cm} P{1.94cm} P{1.94cm}}
        \toprule
\textbf{Data code}      
& \textbf{Description}   
& \textbf{Purpose} 
& \textbf{Detection}
& \textbf{Control}
& \textbf{Uncoloured} \\\midrule
RCI
& Random Generated Image Data      
& Check sampling method isn't introducing parity violation
& $49.8\pm0.3\%$ 
& $50.3\pm0.3\%$ 
& $49.9\pm0.3\%$ \\\hline
HE-CI
& Field of left-handed helices                    
& Check detection of complex parity violation  
& $97.6\pm0.3\%$ 
& $50.5\pm0.3\%$ 
& $62.8\pm0.3\%$ \\\hline
F-HE-CI
& HE-CI, but each helix 50\% chance of being left- or right-handed
& Check method does not detect where there is none
& $51.0\pm0.3\%$ 
& $50.6\pm0.3\%$ 
& $50.5\pm0.3\%$ \\\hline
HE-RCI     
& RCI with 1 left-handed helix inserted into the dataset for every 100 random points                    
& Check detection of complex parity violation  
& $57.5\pm0.3\%$ 
& $50.2\pm0.3\%$ 
& $49.8\pm0.3\%$ \\\hline
F-HE-RCI      
& HE-RCI, but each helix has a  50\% chance of being left- or right-handed 
& Check method of insertion is not causing spurious detection 
& $50.5\pm0.3\%$ 
& $50.4\pm0.3\%$ 
& $50.2\pm0.3\%$ \\\hline
PV-RCI  
& Parity violating groups of 4 points randomly oriented and distributed. Local separation $\ll$ average interpoint distance.  
& Can parity violations be detected where the 2d dataset does not detect.
& $87.2\pm0.1\%$ 
& $49.6\pm0.3\%$ 
& $50.8\pm0.3\%$ \\\hline
PV-RCI-W
&PV-RCI with wider field of view, each image contains $\sim5$ groups of 4.
&Check detection still possible when parity violating groups have very similar pixel colours
& $67.0\pm0.1\%$ 
& $50.2\pm0.3\%$ 
& $50.1\pm0.3\%$ \\\hline
CHE-RCI     
& HE-RCI, but now each helix tapers in, giving it an orientation as well as a handedness               
& Check detection of more complex parity violation  
& $58.9\pm0.3\%$ 
& $49.8\pm0.3\%$ 
& $49.6\pm0.3\%$ \\\hline
F-CHE-RCI      
& CHE-RCI, but each helix has a  50\% chance of being left- or right-handed 
& Check method of insertion is not causing spurious detection 
& $50.5\pm0.3\%$ 
& $50.4\pm0.3\%$ 
& $50.2\pm0.3\%$ \\\hline
HE-RCI-0.1     
& HE-RCI with 10 times fewer helices                 
& Check detection of complex parity violation  
& $54.3\pm0.3\%$ 
& $50.6\pm0.3\%$ 
& $49.9\pm0.3\%$ \\\hline
HE-RCI-0.01     
& HE-RCI with 100 times fewer helices                    
& Check detection of complex parity violation  
& $52.2\pm0.3\%$ 
& $50.1\pm0.3\%$ 
& $49.8\pm0.3\%$ \\\bottomrule
    \end{tabular}
    \caption{A summary of the important colour image based toy datasets. Each dataset is given a code and a description and has its testing purpose outlined. The detection column contains the positive fraction detection; the control column shows the positive fraction for the control group and the uncoloured column shows the detection for the same images coloured monochrome.}
    \label{table:3d_cnn_datacodes_full}
\end{table*}

\begin{table*}
    \begin{tabular}{  p{2.35cm}  p{5.5cm}  p{4.9cm} P{2.0cm} P{2.0cm}}
        \toprule
\textbf{Data code}      
& \textbf{Description}   
& \textbf{Purpose} 
& \textbf{Detection}
& \textbf{Control}\\\midrule
R2d
& Random Generated 2d point Data      
& Check parity violation is not detected where there is none
& $50.41\pm0.10\%$ 
& $50.10\pm0.09\%$ \\\hline
PV-0.1-R2d    
& R2d with two extra points generated near each R2d point, these are generated in a way that all groups of three are of the same chirality. Groups of three are separated on the scale of one tenth the average inter-point distance
& Check detection of parity violating inputs
& $90.77\pm0.06\%$ 
& $49.74\pm0.10\%$ \\\hline
F$x$-PV-0.1-R2d    
& PV-0.1-R2d with 50\% of the extra points inserted with their relative $x$ coordinate to the original point inverted
& Check method of generating points is not injecting parity violation
& $49.66\pm0.10\%$ 
& $49.98\pm0.10\%$ \\\hline
F$y$-PV-0.1-R2d    
& F$x$-PV-0.1-R2d  flipped in $y$ instead of $x$
& Check method of generating points is not injecting parity violation
& $49.98\pm0.10\%$ 
& $49.57\pm0.10\%$ \\\hline
PV-0.1-R2d-5\%  
& PV-0.1-R2d, but only 5\% of inputs generated in this way. The rest are random
& Check fraction of inputs needed to detect
& $52.46\pm0.10\%$ 
& $50.0\pm0.10\%$ \\\hline
PV-0.1-R2d-1\%  
&PV-0.1-R2d, but only 1\% of inputs generated in this way. The rest are random
& Check fraction of inputs needed to detect
& $50.96\pm0.10\%$ 
& $50.0\pm0.10\%$ \\\hline
S-PV-0.1-R2d
&PV-0.1-R2d, but after generating parity violating groups all points shuffled together and new groups drawn at random
&Can parity violations be detected when they are visibly present, and groups sampled randomly from the dataset
& $49.70\pm0.10\%$ 
& $49.68\pm0.10\%$ \\\hline
C-PV-0.1-R2d
&S-PV-0.1-R2d, with a distance constraint when generating groups from the shuffled data   
& Can parity violations become detectable by constraining the scale
& $65.10\pm0.04\%$ 
& $49.85\pm0.04\%$ \\\hline
C-R2d
&R2d with groups constructed after constraining the scale
&Check action of constraining scale is not causing spurious detection
& $50.34\pm0.09\%$ 
& $50.12\pm0.10\%$ \\\hline
C-PV-1-R2d  
&PV-0.1-R2d, but scale of parity violations now same scale as average inter-point distance. No longer visibly parity violating. Groups drawn with scale constraint
& Check parity violations detectable on more realistic scales with scale constraint
& $55.12\pm0.10\%$ 
& $49.63\pm0.10\%$ \\\hline
C-R2d-LH5
&Generate groups of three from R2d with distance in range 1-5 x average inter-point distance. Remove groups of three where $\theta_{12} < \theta_{13}$. Reshuffle and drawn new groups on same scale
&Can different methods of generating parity violations be detected 
& $52.46\pm0.10\%$ 
& $50.07\pm0.10\%$ \\\hline
    \end{tabular}
    \caption{2D toy datasets. Each dataset is given a code and a description and has its testing purpose outlined. The detection column contains the positive fraction detection, and the control column shows the positive fraction for the control group for each test.}
    \label{table:2d_ang_datacodes_full}
\end{table*}

\begin{table*}
    \begin{tabular}{  p{2.35cm}  p{5.5cm}  p{4.9cm} P{2.0cm} P{2.0cm}}
        \toprule
\textbf{Data code}      
& \textbf{Description}   
& \textbf{Purpose} 
& \textbf{Detection}
& \textbf{Control}\\\midrule
R3
& Random Generated 3d point Data      
& Check parity violation is not detected where there is none
& $49.90\pm0.11\%$ 
& $50.11\pm0.10\%$ \\\hline
PV-0.1-R3    
& R3 with three extra points generated near each R3 point, these are generated in a way that all groups of four are of the same chirality. Groups of four are separated on the scale of one tenth the average inter point distance
& Check detection of parity violating inputs
& $87.59\pm0.11\%$ 
& $49.69\pm0.07\%$ \\\hline
F-PV-0.1-R3d    
& PV-0.1-R3d with 50\% of the extra points inserted with their relative x coordinate to the original point inverted
& Check method of generating points is not injecting parity violation
& $50.17\pm0.11\%$ 
& $50.21\pm0.10\%$ \\\hline
PV-0.1-R3-5\%  
& PV-0.1-R3, but only 5\% of inputs generated in this way. The rest are random
& Check fraction of inputs needed to detect
& $53.67\pm0.07\%$ 
& $50.01\pm0.08\%$ \\\hline
PV-0.1-R3-1\%  
&PV-0.1-R3, but only 1\% of inputs generated in this way. The rest are random
& Check fraction of inputs needed to detect
& $49.37\pm0.08\%$ 
& $50.05\pm0.07\%$ \\\hline
S-PV-0.1-R3
&PV-0.1-R3, but after generating parity violating groups all points shuffled together and new groups drawn at random
&Can parity violations be detected when they are visibly present, and groups sampled randomly from the dataset
& $50.37\pm0.08\%$ 
& $50.15\pm0.08\%$ \\\hline
C-PV-0.1-R3
&S-PV-0.1-R3, with a distance constraint when generating groups from the shuffled date   
& Can parity violations become detectable by constraining the scale
& $58.73\pm0.08\%$ 
& $49.87\pm0.08\%$ \\\hline
C-R3
&R3 with groups constructed after constraining the scale
&Check action of constraining scale is not causing spurious detection
& $49.87\pm0.04\%$ 
& $50.32\pm0.07\%$ \\\hline
C-PV-1-R3  
&PV-0.1-R3, but scale of parity violations now same scale as average inter-point distance. No longer visibly parity violating. Groups drawn with scale constraint
& Check parity violations detectable on more realistic scales with scale constraint
& $54.37\pm0.05\%$ 
& $49.87\pm0.06\%$ \\\hline
C-R3-LH5
&Generate groups of 4 from R3 with distance in range 1-5x average inter-point distance. Remove groups of 4 where $\theta_{13} < \theta_{14}$. Reshuffle and drawn new groups on same scale
&Can different methods of generating parity violations be detected
& $53.11\pm0.07\%$ 
& $50.64\pm0.06\%$ \\\bottomrule
    \end{tabular}
    \caption{3D toy datasets. Each dataset is given a code and a description and has its testing purpose outlined. The detection column contains the positive fraction detection, and the control column shows the positive fraction for the control group for each test.}
    \label{table:3d_ang_datacodes_full}
\end{table*}

\begin{table}
\centering
    \begin{tabular}{P{2.5cm} P{2.2cm} P{2.2cm}}
        \toprule
\textbf{Data code}       
& \textbf{Detection}
& \textbf{Control}\\\midrule
R3
& $50.11\pm0.07\%$ 
& $50.21\pm0.06\%$ \\\hline
PV-0.1-R3    
& $99.83\pm0.02\%$ 
& $50.3\pm0.31\%$ \\\hline
F-PV-0.1-R3d 
& $50.35\pm0.07\%$ 
& $49.87\pm0.07\%$ \\\hline
PV-0.1-R3-5\%  
& $53.18\pm0.07\%$ 
& $49.78\pm0.07\%$ \\\hline
PV-0.1-R3-1\%  
& $50.43\pm0.07\%$ 
& $49.91\pm0.07\%$ \\\hline
S-PV-0.1-R3
& $50.16\pm0.07\%$ 
& $49.84\pm0.07\%$ \\\hline
C-PV-0.1-R3
& $47.56\pm0.10\%$ 
& $50.11\pm0.11\%$ \\\hline
C-R3
& $49.98\pm0.11\%$ 
& $50.09\pm0.10\%$ \\\hline
C-PV-1-R3  
& $47.36\pm0.07\%$ 
& $49.91\pm0.07\%$ \\\hline
C-R3-LH5
& $48.12\pm0.07\%$ 
& $50.04\pm0.07\%$ \\\bottomrule
    \end{tabular}
    \caption{Results for the vector input method on the 3d toy datasets, as defined in \cref{table:3d_ang_datacodes_full} The detection column contains the positive fraction detection, and the control column shows the positive fraction for the control group for each test. }
    \label{table:3d_vec_datacodes_full}
\end{table}
\section{Additional tests on networks}\label{appendix:2d_toy_test}

\subsection{2D angles: n=4,5}
As with the method using angles from groups of 3, the methods using 4 and 5 points per group were also tested. These were tested using the same tests as the n=3 case. The results are given in \cref{table:2d_4_datacodes} and \cref{table:2d_5_datacodes}. The tests demonstrate that these methods are equally suitable.
\begin{table}[!htbp]
    \centering
    \begin{tabular}{ c c c}
        \toprule
\textbf{Data code}
& \textbf{Detection}
& \textbf{Control}\\\midrule
R2d
& $50.22\pm1.0\%$ 
& $49.87\pm1.0\%$ \\\hline
PV-0.1-R2d    
& $98.32\pm0.9\%$ 
& $50.91\pm1.0\%$ \\\hline
PV-0.1-R2d-1\%  
& $52.62\pm1.0\%$ 
& $50.43\pm1.0\%$ \\\hline
PV-0.1-R2d-0.5\% 
& $51.56\pm0.3\%$ 
& $50.52\pm1.0\%$ \\\hline
S-PV-0.1-R2d
& $50.36\pm0.3\%$ 
& $50.22\pm1.0\%$ \\\hline
C-PV-0.1-R2d
& $55.57\pm0.3\%$ 
& $50.22\pm1.0\%$ \\\hline
C-R2d
& $50.61\pm0.3\%$ 
& $50.21\pm1.0\%$ \\\hline
C-PV-1-R2d  
& $53.69\pm0.3\%$ 
& $50.32\pm1.0\%$ \\\hline
C-R2d-LH5
& $51.26\pm0.3\%$ 
& $50.62\pm1.0\%$\\\bottomrule
    \end{tabular}
    \caption{2d toy datasets for n=4. The datacodes are defined in \cref{table:2d_ang_datacodes_full}}
    \label{table:2d_4_datacodes}
\end{table}

\begin{table}
    \centering
    \begin{tabular}{ c c c}
        \toprule
\textbf{Data code}
& \textbf{Detection}
& \textbf{Control}\\\midrule
R2d
& $50.24\pm1.0\%$ 
& $50.17\pm1.0\%$ \\\hline
PV-0.1-R2d    
& $99.49\pm0.9\%$ 
& $50.31\pm1.0\%$ \\\hline
PV-0.1-R2d-1\%  
& $53.32\pm1.0\%$ 
& $50.38\pm1.0\%$ \\\hline
PV-0.1-R2d-0.5\% 
& $51.76\pm0.3\%$ 
& $50.21\pm1.0\%$ \\\hline
S-PV-0.1-R2d
& $50.29\pm0.3\%$ 
& $50.27\pm1.0\%$ \\\hline
C-PV-0.1-R2d
& $54.77\pm0.3\%$ 
& $50.18\pm1.0\%$ \\\hline
C-R2d
& $50.51\pm0.3\%$ 
& $50.43\pm1.0\%$ \\\hline
C-PV-1-R2d  
& $53.72\pm0.3\%$ 
& $50.23\pm1.0\%$ \\\hline
C-R2d-LH5
& $51.07\pm0.3\%$ 
& $50.52\pm1.0\%$\\\bottomrule
    \end{tabular}
    \caption{2d toy datasets for n=5. The datacodes are defined in \cref{table:2d_ang_datacodes_full}}
    \label{table:2d_5_datacodes}
\end{table}

\section{Patches of the sky}\label{appendix:binned_sky}
\begin{figure}
    \centering
    \includegraphics[width=0.9\linewidth]{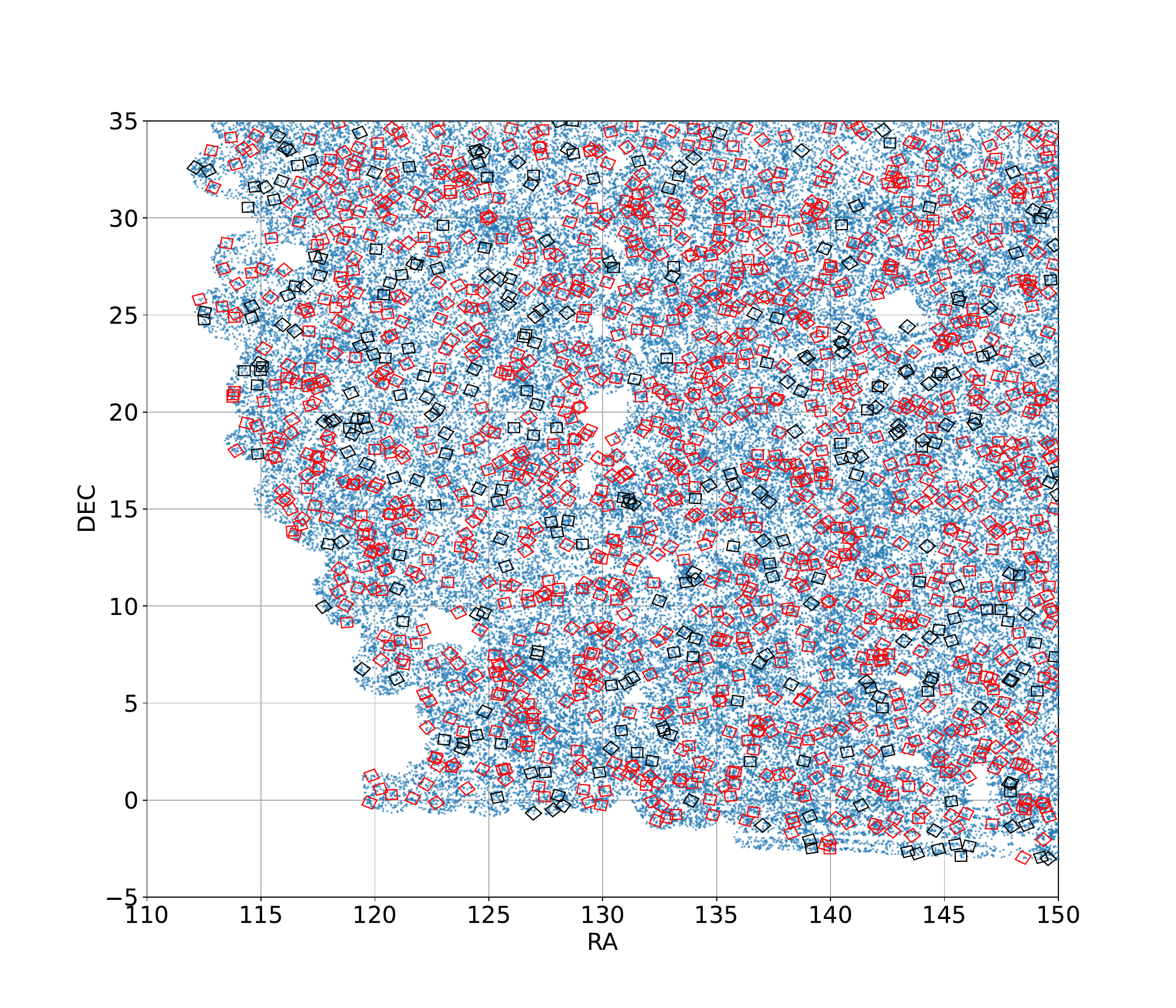}
    \caption{The image samples drawn from the first region for localised ananlysis.}
    \label{fig:Region_1_demo}
\end{figure}

In order to confirm the arguments which propose that the artifacts are having no significant impact, the CMASS NGC was split into eight regions. And image datasets compiled from each separately, the full results follow in \cref{table:regions_results}, where we see that no significant difference occurs between the regions. Suggesting that the artifacts are indeed having no impact. For these tests, the same size patches of the sky were used, and each region had 3,000 images generated from it. The image distribution for the first region is shown in \cref{fig:Region_1_demo}

\begin{table}
    \centering
    \begin{tabular}{ c   c   c}
        \toprule
\textbf{Data code}
& \textbf{Detection}
& \textbf{Control}\\\midrule
R1
& $49.96\pm2.0\%$ 
& $53.43\pm2.0\%$ \\\hline
R2  
& $48.95\pm2.0\%$ 
& $51.33\pm2.0\%$ \\\hline
R3
& $50.91\pm2.0\%$ 
& $50.4\pm2.0\%$ \\\hline
R4
& $51.31\pm2.0\%$ 
& $50.92\pm2.0\%$ \\\hline
R5
& $49.65\pm2.0\%$ 
& $52.43\pm2.0\%$ \\\hline
R6
& $52.46\pm2.0\%$ 
& $51.04\pm2.0\%$ \\\hline
R7
& $48.76\pm2.0\%$ 
& $52.93\pm2.0\%$ \\\hline
R8
& $49.96\pm2.0\%$ 
& $53.43\pm2.0\%$ \\\bottomrule
    \end{tabular}
    \caption{Levels of detection in the separate regions, used to investigate the impact of the artifacts.}
    \label{table:regions_results}
\end{table}

\end{document}